\documentclass[11pt,a4paper]{article}
\usepackage{eurosym}
\usepackage{amsthm}
\usepackage{amsfonts}
\usepackage{amsmath,mathrsfs}
\usepackage{amssymb}
\usepackage{graphicx}
\usepackage{color}

\newtheorem{theorem}{Theorem}

\newtheorem{corollary}{Corollary}

\newtheorem{definition}{Definition}

\newtheorem{lemma}{Lemma}

\newtheorem{proposition}{Proposition}
\newtheorem{remark}{Remark}

\newtheorem{lettertheorem}{Theorem}
 
\newtheorem{letterprop}{Proposition}

\definecolor{awesome}{rgb}{1.0, 0.13, 0.32}

\numberwithin{equation}{section}

\begin{document}
\title{Inverse problems for the zeros of the Wigner function}
\author{Lu\'{\i}s Daniel Abreu \\
%EndAName
University of Vienna \and Ulysse Chabaud \\
%EndAName
DIENS, \'Ecole Normale Sup\'erieure, PSL University, CNRS, INRIA (QAT) \and %
Nuno Costa Dias \\
%EndAName
University of Lisbon (GFM) \and Jo\~ao Nuno Prata \\
%EndAName
ISCTE-IUL and \\
University of Lisbon (GFM)}
\maketitle

\begin{abstract}
We study the nodal (zero) set of the Wigner transform, obtaining a rigidity
theorem for the eigenstates of one-dimensional harmonic oscillator. For
every $k \in \mathbb{N}_0$, and $f\in L^{2}(\mathbb{R})$, equality between
the nodal sets of $Wf$ and $Wh_{k}$ determines $f$, up to a constant phase
and an explicitly defined nodal-preserving pseudo-displacement. The proof
requires a result with independent interest, where it is shown that a Wigner
distribution has bounded nodal set if and only if its underlying state is,
up to time-frequency and metaplectic symmetries, a finite Hermite expansion.

The demonstration of the results combines new analytic, geometric and
arithmetic tools: a property of classical convolution, a geometric study of
phase space displacements and a new divisibility theorem for generalized
Laguerre polynomials.

As a by-product, we obtain several results concerning the structure and
admissible geometry of bounded nodal sets of Wigner distributions.
\end{abstract}

%
%-------%
% TITLE %
% TITLE %
%-------%
%------------------------------------------%
%------------------------------------------%

\section{Introduction}

Rigidity theorems are a recurrent theme in mathematics. The term originates
in the geometric notion of resistance to deformation. It expresses the
principle that apparently partial or qualitative information may determine a
mathematical object completely, up to the natural symmetries of the problem.
In analysis on Euclidean space, there is a remarkable number of rigidity
results involving Gaussian functions. But for a single prescribed Hermite
function, the only known characterizations involve global information, such
as their classical algebraic or spectral conditions, or the spectral data of
the Harmonic oscillator.

Early rigidity results in Fourier analysis, are provided by the Gaussian
characterization of the equality cases in Heisenberg--Kennard's \cite%
{Kennard} and by Hardy's uncertainty principles \cite{Hardy}. A very deep
result is the characterization of the Gaussian extremizers in Beckner's
sharp inequalities \cite{Beckner}, together with Lieb's proof of uniqueness
up to natural isometries \cite{LiebRigidity}. A second class of Fourier
rigidity theorems arises from polynomially relaxed forms of Hardy's
uncertainty principle \cite{Hardy}. Whereas the critical Hardy theorem
singles out the Gaussian, allowing polynomial factors in the critical decay
estimates leads to finite linear combinations of Hermite functions. This
principle was refined by Bonami, Demange and Jaming \cite{Bonami}, who
characterized Gaussian growth with a polynomial loss of a function and its
Fourier transform. We were unable to find a Fourier rigidity theorem
characterizing a prescribed Hermite function.

In the phase space we have a similar picture. Hudson's theorem \cite%
{Hudson,Janssen2}, which will be the first main reference for our
discussion, states that, among pure states, only generalized Gaussians have
non-negative Wigner transform. Lieb \cite{Lieb} and Carlen \cite{Carlen}
established the rigidity of Gaussians as minimizers of Wehrl entropy, while
Lieb \cite{LiebWigner} proved this rigidity for all $p$-norms of the Wigner
transform, a result that contains Hudson's theorem as a special case. More
recently, Nicola and Tilli \cite{NicolaTilli2022} proved that the optimal
Rayleigh quotient of Daubechies localization operators \cite{Daubechies1988}
is completely rigid for the circular shape of the localization domain and
the corresponding Gaussian maximizer. This led to a sharp geometric
Gaussianity deficit measure, presented as a stability result in \cite%
{GuerraGomesTilliRamos}. The Hardy theorems of Gr\"{o}chenig-Zimmerman \cite%
{GroHardy} and de Gosson-Luef \cite{Luef} provide a phase space Hardy-type
rigidity, while Bonami--Demange--Jaming \cite{Bonami} give more flexible
polynomial-Gaussian characterizations. We conclude this literature review by
observing that Janssen's \cite{JanssenRadial} results imply that, if $Wf$ is
radial, then $f=h_{k}$ for some $k\in \mathbb{N}$. Again, we were unable to
find a phase space rigidity theorem characterizing a prescribed Hermite
function. The purpose of the present work is to show that substantially
weaker phase-space information---\emph{the nodal set of a particular $Wh_k$}%
---determines the corresponding Hermite function $h_k$, modulo a phase and a
certain pseudo-displacement symmetry to be defined below. Our results are close in spirit to various old and new results in harmonic analysis \cite{Benedicks,Kulikov,Paley,Radchenko,Ramos}.

Before stating our main result, we need to fix some definitions and
notation. Given $f,g\in L^{2}(\mathbb{R})$ the cross-Wigner function is
given by \cite{Gosson,Lerner,Lions,Wigner}: 
\begin{equation}
W(f,g)(x,p)=\frac{1}{2\pi \hbar }\int_{\mathbb{R}}f(x+\tau /2)\overline{%
g(x-\tau /2)}e^{-\frac{i}{\hbar }p\tau }d\tau \text{,}  \label{eqrevision1}
\end{equation}%
where $h=2\pi \hbar $ is Planck's constant and $x,p$ are interpreted as the
position and momentum variables, respectively. In time-frequency analysis, $%
x $ is the time and $p$ the frequency variable and typically $\hbar =\frac{1%
}{2\pi }$. We denote by $z=x+ip\equiv (x,p)\in \mathbb{R}^{2}$ a point in
the phase-space or time-frequency plane $\mathbb{R}^{2}\equiv \mathbb{C}$.
The Wigner distribution of $f$ is $Wf(z)\equiv W(f,f)(z)$: 
\begin{equation}
Wf(x,p)=\frac{1}{2\pi \hbar }\int_{\mathbb{R}}f(x+\tau /2)\overline{f(x-\tau
/2)}e^{-\frac{i}{\hbar }p\tau }d\tau \text{.}  \label{eqrevision2}
\end{equation}%
In this work we shall always assume that $Wf$ is normalized: $\Vert f\Vert
_{2}=1$, which implies that%
\begin{equation}
\int_{\mathbb{R}^{2}}Wf(x,p)dxdp=1\text{.}  \label{Normalization}
\end{equation}%
The \emph{nodal set }of $Wf$ is: \emph{\ } 
\begin{equation}
\mathcal{N}(Wf):=\left\{ z\in \mathbb{R}^{2}:Wf(z)=0\right\} \text{.}
\label{eqrevision7}
\end{equation}%
This set is consistent as the Wigner distribution is well-defined pointwise
(in fact it is uniformly continuous \cite{Grochenig}). To state our main
results we will need the usual notation for a time-frequency shift, 
\begin{equation*}
\left( \pi (z_{0})f\right) (x)=e^{\frac{i}{\hbar }p_{0}x}f(x-x_{0})~,\hspace{%
0.3cm}z_{0}=(x_{0},p_{0})\text{.}
\end{equation*}%
For $z_{0}=(x_{0},p_{0})\in \mathbb{R}^{2}$, let $E_{z_{0}}:=\pi
(z_{0})D_{-\alpha }$, where $D_{-\alpha }$ is the pseudo-displacement
operator defined in (\ref{eqpseudoDisp16}) and $\alpha =\tfrac{1}{\hbar }%
(-x_{0}+ip_{0})$. The operator $E_{z_{0}}$ moves the centre of the Gaussian
envelope of the corresponding Wigner function while leaving the polynomial
factor-and therefore its zero set-unchanged (see also Definition \ref%
{DefinitionOperatorE}). With these notations we are in a position of stating
the main result of this paper.

\begin{theorem}
\label{MainTheorem}For every $k\in \mathbb{N}$ and $f\in L^{2}(\mathbb{R})$, 
\begin{equation}
\mathcal{N}(Wh_{k})=\mathcal{N}(Wf)  \label{NodalContainer}
\end{equation}%
if and only if 
\begin{equation}
Wf(z)=cWh_{k}(z)e^{-\frac{z_{0}}{\hbar }\cdot (z_{0}-2z)}~,
\label{eqMainTheo1}
\end{equation}%
for a normalization constant $c\in \mathbb{R}^{+}$ determined by (\ref%
{Normalization})\ and $z_{0}\in \mathbb{R}^{2}$. In particular, (\ref%
{NodalContainer}) is equivalent to%
\begin{equation}
f=e^{i\theta }\frac{E_{z_{0}}h_{k}}{\left\Vert E_{z_{0}}h_{k}\right\Vert _{2}%
}  \label{eqMainTheo2}
\end{equation}%
for some $\theta \in \left[ 0,2\pi \right] $. If we assume that $Wf(z)$ has
bounded nodal set, then (\ref{NodalContainer}) may be replaced by the weaker
condition $\mathcal{N}(Wh_{k})\subseteq \mathcal{N}(Wf)$.
\end{theorem}

A comparison between Theorem \ref{MainTheorem} for $k\geq 1$ and Hudson's
theorem for $k=0$ reveals a \emph{sharp symmetry breaking} in the inverse
nodal problem: the invariance of the case $k=0$ under affine isometries of
the plane is lost for $k\geq 1$. Precisely, Hudson's theorem tells that $%
\mathcal{N}(Wf)=\emptyset $ forces $f$ to be a generalized Gaussian:%
\begin{equation}
f=e^{i\theta }\pi (z_{0})\mu (S^{-1})h_{0}\text{, \ \ }z_{0}\in \mathbb{R}%
^{2}\text{,\ \ \ \ }S\in Sp(2,\mathbb{R})\text{,}
\label{eqfunctionHudsonTheorem}
\end{equation}%
where the notation $\mu (S)$ is a metaplectic operator associated with $S$.
The reduction from $Sp(2,\mathbb{R})$ in (\ref{eqfunctionHudsonTheorem})\ to
its maximal compact subgroup $SO(2)$ in (\ref{eqMainTheo2}) is not a
limitation: this symmetry breaking is determined by the symmetries of the
nodal sets of $Wh_{k}$ when $k=0$ and $k\neq 0$ (see Remark \ref%
{SymmetryBreaking}) for a more detailed discussion).

We also observe that a characterization of a function in terms of its zeros
does not exist even in the case of the Bargmann transform \cite{Bargmann},
since there are many functions with $k$ zeros at the origin. This might come
as a surprise, specially if we take into account that the Wigner transform
lacks almost the whole structure required in an approach using established
mathematical tools. The Wigner distribution is not a linear object and
possesses no holomorphic structure (it is a real function of a complex
variable), therefore severely constraining any attempt of treating it with
classical complex analysis, or conventional functional analysis. This lack
of structure, which this work reveals to be only apparent, also explains
why, despite the extensive literature on Wigner transforms themselves, the
study of Wigner nodal sets has received comparatively little attention. To
the best of our knowledge, systematic results on non-trivial zero sets of
Wigner transforms are scarce (one of the few works on the topic is the
relatively recent paper \cite{Jaming1}, where zero-free cross-Wigner
distributions are investigated). To overcome the obstructions required for
the proof of Theorem \ref{MainTheorem} we resort to a combination of Fourier
and time-frequency analysis, symplectic covariance, a trigonometric
polynomial decomposition, and the Sylvester--Schur theory of prime divisors.
Among the intermediate results obtained, two of them have interest on their
own sake, and will now be presented as our second and third main theorems.

Our second main result is a different rigidity theorem, characterizing
Wigner functions with a bounded nodal set.

\begin{theorem}
\label{TheoremSmootDecay}Let $f\in L^{2}(\mathbb{R})$. Then $Wf(z)$ has
bounded nodal set if and only if 
\begin{equation}
f=\pi (z_{1})\mu (S^{-1})\sum_{n=0}^{N}b_{n}h_{n}\text{ , \ \ \ \ \ \ }S\in
Sp(2,\mathbb{R})  \label{eqTheoremEllipse1}
\end{equation}%
for some $N\in \mathbb{N}_{0}$, $b_{0},\cdots ,b_{N}\in \mathbb{C}$, $%
z_{1}\in \mathbb{R}^{2}$, and a symplectic matrix $S$, where $h_{n}$ denotes
the $n$-th Hermite function.
\end{theorem}

We define the \emph{Hermite rank} of $f$ (or simply the \emph{rank}) to be
the largest $N$ such that $b_{n}\neq 0$. By Theorem \ref{TheoremSmootDecay},
boundedness of the nodal set is equivalent to finite \emph{Hermite rank} $N$%
. The integer $N$ is also the number of zeros of the corresponding Bargmann
transform, counting multiplicities, and is closely related to the stellar
rank \cite{Ulyss} used in quantum information.

The proof of Theorem \ref{MainTheorem} uses Theorem \ref{TheoremSmootDecay}
to write the Wigner transform $Wf$, with $f$ of the form (\ref{eqTheoremEllipse1}), as a sum of products of polynomials in $(z,\overline{z})
$, whose coefficients are Laguerre polynomials of the form $L_{n}^{(m)}$.
Forcing this sum to vanish on the (circular) zeros of $Wh_{k}$ leads to a
trigonometric polynomial structure, which is then used to reduce the
rigidity of the zeros of $Wh_{k}$ to the rigidity of the roots of the
Laguerre polynomial $L_{k}$. To exclude the possibility that a different
finite Hermite expansion has all of these circles among its zeros, the
following divisibility theorem is finally used to complete the proof of
Theorem \ref{MainTheorem} (after this it is only required to remove the
centered assumption, we will introduce the pseudo-displacement operator in
section \ref{SubsectionPseudoDispOp}). 

\begin{theorem}
\label{TheoremLaguerreDivisibility} Let $n,k \in \mathbb{N}$ and $m \in 
\mathbb{N}_0$. If 
\begin{equation}
L_k(x) |L_n^{(m)}(x)  \label{eqTheoremLaguerreDivisibility1}
\end{equation}
in $\mathbb{C} \left[x \right]$, then 
\begin{equation}
n=k~, \hspace{0.5 cm} m=0~.  \label{eqTheoremLaguerreDivisibility2}
\end{equation}
Equivalently, 
\begin{equation}
L_k = L_n^{(m)}~.  \label{eqTheoremLaguerreDivisibility3}
\end{equation}
\end{theorem}

This result is of independent arithmetic interest. It settles, for integer
exponents, a divisibility problem for generalized Laguerre polynomials,
whose investigation was initiated by I. Schur \cite{Schur1,Schur2} in 1929
as is still of current research interest in pure mathematics \cite%
{Filaseta1,Filaseta3,Shorey2,Shorey1}. The theorem is precisely the rigidity
statement needed to turn the geometric information carried by the Laguerre
zeros into uniqueness of the underlying Wigner distribution.

The combination of Theorems \ref{TheoremSmootDecay}--\ref%
{TheoremLaguerreDivisibility} suggests a broader principle. The zero set of
a Wigner distribution is, in general, highly incomplete information.
Nevertheless, once the nodal geometry is subject to an asymptotic constraint
(bounded nodal set), it can become an unexpectedly rigid invariant of the
state. The results therefore provide a first framework for an \emph{inverse
theory based on Wigner nodal sets}, in which geometric information in phase
space is converted into algebraic information about the underlying state.

There are several indications that this phenomenon is not restricted to the
particular global condition considered here. We prove, for example,
restrictions on the possible geometry of bounded nodal sets, including the
occurrence of circles, ellipses, line segments and isolated zeros, and show
that certain apparently local pieces of nodal geometry already force global
algebraic constraints. In particular, the presence of a circular arc can
impose the same rigidity as the corresponding complete circle. These results
point toward inverse questions in which one seeks to determine a state from
only partial observations of its nodal set. Our results thus place the
geometry of Wigner zeros in a broader analytic context: \emph{global
information about the zero set can force global algebraic structure of the
state, while sufficiently rich nodal geometry can determine the phase-space
distribution itself.} We view this as the beginning of a classification
theory of Wigner nodal sets, which, according to the results in this paper,
share some properties with those of entire functions.

From a time-frequency analysis viewpoint, the problem is closely related to
the structure of quadratic time-frequency representations and their
interaction with finite Hermite expansions. From the perspective of quantum
mechanics and quantum information, the results provide a geometric
characterization of a distinguished class of non-Gaussian states and suggest
that nodal data may serve as a form of state characterization. Moreover,
negativity is usual regarded as a measure of quantum features like
entanglement \cite{Dahl,Kenfack}. More broadly, they raise quantitative and
higher-dimensional questions: how much of a nodal set is required for
reconstruction, whether analogous rigidity holds for multimode Wigner
distributions, and whether approximate nodal information implies approximate
finite Hermite rank. A possible approach to these problem could start by
obtaining a Hermite family version of the stability properties proved in 
\cite{GuerraGomesTilliRamos} for the Gaussianity deficit.

\section{Outline of the paper}

The paper is outlined as follows. In the next section we describe the
strategy of the proof of the main results and we gather various additional
results on the rank of the Hermite expansion or the geometry of the nodal
set, which require Theorem \ref{TheoremSmootDecay} in their proofs. Then we
collect the notational remarks, the important explicit formulas of the
Wigner transforms of Hermite functions and some remarks about the connection
of our setting with the theory of the polyanalytic functions (which is not
used in the proofs of any of our results) (section \ref{Section2}). Next we
present various known properties of Wigner functions which will be used in
the proofs (section \ref{Section3}). In section \ref%
{SectionLaguerrePolynomials} we recall several properties of Laguerre
polynomials. The core of the paper is section \ref{Section4}, where we
compile the proofs of our main results.

\subsection{Strategy of the proof}

Here we briefly discuss the strategy of the proofs of Theorems \ref%
{TheoremSmootDecay} and \ref{MainTheorem}.

In the proof of Hudson's theorem \cite{Grochenig,Hudson}, one computes the
Husimi function of $f$. This amounts to the convolution of the Gaussian $%
Wh_{0}$ with $Wf$. If $Wf$ has no zeros, then the Husimi function has no
zeros. On the other hand the Husimi function can be roughly seen as the
modulus squared of the Bargmann transform of $f$. Since the Bargmann
transform is an entire function with order of growth two, then if has no
zeros, it must be a generalized Gaussian by the Weierstrass factorization
theorem. The arguments used to prove Theorem \ref{TheoremSmootDecay}
departure from the same basic principles as \cite{Grochenig,Hudson}.
Actually, the proof shares a step with the proof of \cite[Proposition 6.2]%
{Jaming1}, where it is concluded that if the short-time Fourier transform
(STFT) of a function $f$ \ (a small modification of the cross-Wigner
transform (\ref{eqrevision1})), windowed by a function $g$ of the form (\ref%
{eqTheoremEllipse1}), is void of zeros (therefore with a bounded nodal set),
then $f$ must be itself of the form (\ref{eqTheoremEllipse1}) (but not
necessarily of the same `rank' $N$ as $g$). However, in our quadratic case,
we are not assuming a priori that $f$ is of the form (\ref{eqTheoremEllipse1}%
). Thus, the resulting transform is not necessarily polyentire and Balk's
theorem (\cite{Balk}, \cite[Theorem 6.1]{Jaming1}) cannot assure the
polynomial structure. To circumvent this obstruction, we need to show that
boundedness of $\mathcal{N}(Wf)$ is preserved by convolution with $Wh_{0}(z)=%
\frac{1}{\pi \hbar }e^{-\frac{|z|^{2}}{\hbar }}$. It turns out that this is
a general property of convolution, but we could not find it in the
literature. Thus, it is stated below and proved in section \ref%
{SubsectionConvolution}. It may have independent interest. We will only need
the dimension $n=2$, but since the proof is the same for all $n\geq 2$ and,
with some minor changes, for $n=1$, we state and prove it for arbitrary
dimension.

\begin{theorem}
\label{TheoremConvolution} Let $f\in C(\mathbb{R}^{n})\cap L^{\infty }(%
\mathbb{R}^{n})$ be a real valued function such that its nodal set is
bounded. Then the nodal set of the convolution $f\star g$ with the Gaussian $%
g(x)=e^{-|x|^{2}}$ is also bounded.
\end{theorem}

This transfers geometric information on the nodal set of $Wf$ to the Husimi
function and hence to the Bargmann transform. The latter is entire of order
at most two, so Weierstrass factorization converts boundedness of its zero
set into a polynomial-exponential representation. Pulling this
representation back to $L^{2}(\mathbb{R})$ yields the Gaussian-polynomial
structure of $f$.

To prove Theorem \ref{MainTheorem} our starting point is Theorem \ref%
{TheoremSmootDecay}. We then determine the form of the function $f$ and its
Wigner function if the nodal set contains an ellipse.

\begin{theorem}
\label{TheoremEllipse} Let $Wf(z)$ be a Wigner function with a bounded nodal
set. If $\mathcal{N}(Wf)$ contains some ellipse $(z-z_{0})\cdot M(z-z_{0})=1$
(with $z_{0}\in \mathbb{R}^{2}$ and $M$ a $2\times 2$ real, symmetric,
positive-definite matrix), then there exist $N\in \mathbb{N}$, $%
(c_{0},c_{1},\cdots ,c_{N})\in \mathbb{C}^{N+1}$, and $z_{1}\in \mathbb{R}%
^{2}$, such that $f$ is of the form (\ref{eqTheoremEllipse1}), where $S$ is
a (non-unique) symplectic matrix such that 
\begin{equation}
M=\sqrt{\det M}~S^{T}S~.  \label{eqTheoremEllipse2}
\end{equation}%
The associated Wigner distribution is: 
\begin{equation}
Wf(z)=\sum_{n=0}^{N}\sum_{m=0}^{N}c_{n}\overline{c_{m}}%
W(h_{n},h_{m})(S(z-z_{1}))\text{,}  \label{eqTheoremEllipse3}
\end{equation}%
where $z_{0}$ is the center of the ellipse and $z_{1}$ the center of the
Gaussian.
\end{theorem}

In particular, if the ellipse is a circle, we can simply choose $S=I$. In
the sequel we will simplify our phrasing by saying that $Wf(z)$\emph{\
admits a circular zero of radius} $R\in \mathbb{R}^{+}$ if a circle $\frac{%
2|z-z_{0}|^{2}}{\hbar }=R$ is contained in $\mathcal{N}(Wf)$ for some $%
z_{0}\in \mathbb{R}^{2}$. In this situation, we can omit $S$ in the
representation (\ref{eqTheoremEllipse3}). To obtain a more precise form of $%
Wf$ we first observe that the center $z_{1}$ of the Gaussian does not
necessarily coincide with the center $z_{0}$ of the radial zero. To overcome
the resulting technical challenge, we resort to an operator - the
pseudo-displacement operator - which upon action on a function which is a
polynomial times a Gaussian, shifts the polynomial while leaving the
Gaussian part unchanged (see section \ref{SubsectionPseudoDispOp} below).
This (together with a global time-frequency shift) shows that we can assume $%
Wf$ to be of the form $Wf(z)=KWg(z)e^{-\frac{1}{\hbar }(z_{1}^{2}-2z_{1}z)}$%
, where 
\begin{equation*}
Wg(z)=\sum_{n=0}^{N}\sum_{m=0}^{N}c_{n}\overline{c_{m}}W(h_{n},h_{m})(z)
\end{equation*}%
has a circular zero centered at the origin. This will simplify all
calculations substantially.

Then we assume that $\mathcal{N}(Wh_k) \subset \mathcal{N}(Wf)$. That means
that $Wg$ has all the circular zeros $\frac{2|z|^2}{\hbar}=R$, where $R=
\rho_j^{(k)}$, $j=1, \cdots, k$ are the $k$ positive roots of $L_k$. Using
Theorem \ref{TheoremLaguerreDivisibility} we can then prove that this is
possible iff $Wg=Wh_k$, which proves Theorem \ref{MainTheorem}.

Finally, in the proof of Theorem \ref{TheoremLaguerreDivisibility} we
consider different techniques for the ranges $k\leq n<2k$ and $n\geq 2k$. In
the former range, we consider the ordinary differential equations for the
two polynomials as well as the divisibility and the degree of the
polynomials to conclude that they have to be equal. In the latter range, we
form a congruence tower and use a theorem by Sylvester and Schur for prime
divisors, to show that $L_{k}$ cannot divide $L_{n}^{(m)}$.

\begin{remark}
\label{SymmetryBreaking} Before we conclude let us make some remarks on the
symmetry breaking of the inverse problem for the nodal sets when $k=0$ and $%
k \neq 0$.

To place the two results in a comparable form, we make the following
observation. We can rewrite eqs.(\ref{eqMainTheo2}, \ref%
{eqfunctionHudsonTheorem}) in a unified form: 
\begin{equation}
f= e^{i\theta} \frac{E_{z_0}\mu (S^{-1}) h_k}{\|E_{z_0} h_k\|_2}~,
\label{eqRemarkSymmetryBreaking1}
\end{equation}
for all $k \in \mathbb{N}_0$.

Indeed, if we take Hudson's theorem and choose $k=0$, then $E_{z_0}$ acting
on a generalized Gaussian is tantamount to a displacement (see section \ref%
{subsectionPseudoDispOp}): $E_{z_0}\mu(S^{-1}) h_0=M \pi(z_0) \mu(S^{-1})h_0$%
, where $M$ is some constant, and we recover \ref{eqfunctionHudsonTheorem}.

Alternatively, if $k \geq 1$, then from Theorem \ref{TheoremEllipse}, $S$
must also be an orthogonal matrix. Hence, $h_k$ is an eigenstate of $%
\mu(S^{-1})$ and we recover (\ref{eqMainTheo2}).

The symmetry breaking when moving from $k=0$ to $k\geq 1$, is easy to
describe geometrically: for $k=0$, the nodal set of the Gaussian $h_{0}$ is
empty and, trivially, contains no nodal set geometry capable of detecting a
nonorthogonal symplectic transformation $S\in Sp(2,\mathbb{R})$; such $S$
sends circles to ellipses, therefore keeping the void nodal set of $Wh_{0}$
invariant. This symmetry is broken for $k\geq 1$: the nodal set $\mathcal{N}%
(Wh_{k})$ contains the $k$ concentric circular zeros arising from the
Laguerre factor (see (\ref{eqWigner24})) and it is deformed by a
nonorthogonal symplectic transformation $S\notin SO(2)$.

It is possible to exactly identify the group whose action is lost in this
symmetry breaking pattern. Since a nonorthogonal symplectic transformation $%
S\notin SO(2)$\ deforms these circles into ellipses, by requiring the
preservation of the circular nodal set $\mathcal{N}(Wh_{k})$ one is reducing
the full symplectic stabilizer group $Sp(2,\mathbb{R})$ to its maximal
compact subgroup $SO(2)$. Consequently, the broken symmetry is precisely the
symmetric space%
\begin{equation*}
\frac{Sp(2,\mathbb{R})}{SO(2)}\text{,}
\end{equation*}%
which, by Iwasawa decomposition, can be identified with the action of the
affine group $ax+b$ in $\mathbb{R}^{2}$. On phase space, $a$ acts as the
squeezing map $(q,p)\rightarrow (\sqrt{a}q,p/\sqrt{a})$, while $b$ is the
linear shear $(q,p)\rightarrow (q+bp,p)$.\ These are exactly the actions
that deform the nodal set of $\mathcal{N}(Wh_{k})$, while trivially keeping
the void $\mathcal{N}(Wh_{0})$ of Hudson's theorem invariant and
characterize the group action lost for $k\geq 1$. \emph{In both cases, the
conclusion is optimal with respect to the information carried by the Wigner
nodal set.}
\end{remark}

\subsection{Remark on the previous status of Theorem \protect\ref%
{TheoremLaguerreDivisibility}}

Some special cases of Theorem \ref{TheoremLaguerreDivisibility} have been
previously proved, in the broader context of the divisibility problem for
Laguerre polynomials over the rationals. Research on this was initiated in
1929 by Schur, who started to address systematically the problem of the
factorization of generalized Laguerre polynomials over the rationals, and
proved \cite{Schur1,Schur2} that $L_{n}^{(0)}(x)$, $L_{n}^{(1)}(x)$ and $%
L_{n}^{(-n-1)}(x)$ do not factorize over the rationals. Filaseta and Lam
proved \cite{Filaseta1} that, for fixed rational $\alpha $, there are only
finitely many Laguerre polynomials which factorize over the rationals, using
a method first developed for the Bessel polynomials \cite{Filaseta2}.
Laishram and Shorey \cite[Theorem 1]{Shorey1} proved that for $0\leq \alpha
\leq 50$, $L_{n}^{(\alpha )}(x)$ is irreducible over the rationals except
for $8$ explicit cases of $(n,\alpha )$. Moreover, in \cite[Theorem 2]%
{Shorey1}, they proved that, given $1\leq k\leq \frac{n}{2}$ and $0\leq
\alpha \leq 5k$, $L_{n}^{(\alpha )}$ has no factor of degree $k$ except when 
$k=1$ and $(n,\alpha )\in \left\{ (2,2),(4,5)\right\} $. In \cite{Shorey2}
Fuchs and Shorey give various other conditions for a generalized Laguerre
polynomials to have factors of a given degree $k$. Finally, Filaseta, Finch
and Leidy \cite{Filaseta3} reviewed various results in this context. Besides
the results already stated above, they considered other cases where $\alpha $
is negative and the case $\alpha =n$. In the latter case, $L_{n}^{(n)}$ only
factorizes over the rationals for $n=2$.

\subsection{Further results on the nodal set}

\subsubsection{Structure theorems}

The first immediate consequence of Theorem \ref{TheoremSmootDecay} is the
fact that \emph{any domain in }$\mathbb{R}^{2}$\emph{\ is a uniqueness set
for Wigner functions with bounded nodal set,} a property reminiscent of
analytic functions.

\begin{corollary}
\label{CorollaryUniquenessSet} Consider two Wigner functions $Wf$ and $Wg$
with bounded nodal sets. If $Wf$ and $Wg$ coincide on some domain $D$ (open,
connected set), then $Wf=Wg$ everywhere in $\mathbb{R}^{2}$.
\end{corollary}

\begin{remark}
\label{Uniquenesssets} At this point, we should add a note of caution. In
harmonic analysis, the problem of determining uniqueness sets and vanishing
sets for a function and its Fourier (or any linear) transform is basically
the same. This can be stated as follows. Two sets $\Omega ,~\widehat{\Omega }%
\subset \mathbb{R}^{n}$ are vanishing sets for the Fourier transform if 
\begin{equation*}
f(x)=0~,~\forall x\in \Omega ~\text{ and }~\mathcal{F}_{\hbar
}f(p)=0~,~\forall p\in \widehat{\Omega }~\Rightarrow f\equiv 0~.
\end{equation*}%
But these sets are also called uniqueness sets because we have the
straightforward equivalent statement: 
\begin{equation*}
f(x)=g(x)~,~\forall x\in \Omega ~\text{ and }~\mathcal{F}_{\hbar }f(p)=%
\mathcal{F}_{\hbar }g(p)~,~\forall p\in \widehat{\Omega }~\Rightarrow
f\equiv g~.
\end{equation*}%
However, due to the quadratic nature of the Wigner function, we cannot
conclude from 
\begin{equation*}
Wf(z)=0~,~\forall z\in \Sigma \subset \mathbb{R}^{2}~\Rightarrow Wf\equiv 0~
\end{equation*}%
that 
\begin{equation*}
Wf(z)=Wg(z)~,~\forall z\in \Sigma \subset \mathbb{R}^{2}~\Rightarrow
Wf\equiv Wg~\text{,}
\end{equation*}%
since $Wf(z)-Wg(z)\neq W(f-g)(z)$. Nevertheless, it is true that the latter
statement implies the former. The coincidence of uniqueness and vanishing
sets does not hold for Wigner distributions.
\end{remark}

\begin{remark}
\label{RemarkPhaseRetrieval} We remark that Corollary \ref%
{CorollaryUniquenessSet} solves the phase retrieval problem for Wigner
functions, when the nodal set is bounded. \cite{Jaming}:

Suppose that $|Wf(z)|=|Wg(z)|$ for all $z \in \mathbb{R}^2$. When can we
conclude that $f=c g$ for some $c \in \mathbb{C}$ with $|c|=1$?

\vspace{0.2 cm} In the terminology of \cite{Jaming}, Wigner functions
satisfying $|Wf(z)|=|Wg(z)|$ are said to be \textit{Wigner partners}.
Suppose that $Wf$ has bounded nodal set, and $Wf$, $Wg$ are Wigner partners.
Then $Wg$ also has a bounded nodal set, and there exists $R_0 >0$ such that $%
\mathcal{N}(Wf) \cup \mathcal{N}(Wg) \subset B_{R_0}$. Hence, we have $Wf(z)
= +Wg(z)$ or $Wf(z) = -Wg(z)$ for all $|z| >R_0$. However, Hardy's theorem
for Wigner distributions \cite{Luef,GroHardy} tells us that both Wigner
functions must be positive for $|z| >R_0$. From Corollary \ref%
{CorollaryUniquenessSet} it follows that $Wf(z)=Wg(z)$ everywhere. By the
inversion of the Wigner transform, we conclude that $f=cg$.
\end{remark}

We now return to the characterization of nodal sets of Wigner functions.

\begin{theorem}
\label{TheoremRadialZeros} Let $Wf(z) $ be a Wigner function with bounded
nodal set. Then all admissible circular zeros of $Wf$ have radii which are
roots of some Laguerre polynomial $L_{n}^{(m)}(x)$, $n\in \mathbb{N}$, $m\in 
\mathbb{N}_{0}$.
\end{theorem}

We proceed with a result on admissible rational circular zeros. Clearly,
Theorem \ref{TheoremRadialZeros} implies that the set of all possible radii
where a Wigner function with bounded nodal set admits a circular zero, is
countable. In the next result we show that, among rational numbers, only
positive integers can be radii of admissible circular zeros.

\begin{theorem}
\label{TheoremRationalRadius} Let $Wf(z)$ be a Wigner function with a
bounded nodal set. Suppose that $Wf(z)$ has a circular zero, centered at $%
z_0 \in \mathbb{R}^2$, with radius $p\in \mathbb{Q}$. Then, this is possible
only if $p\in \mathbb{N}$. In this case, if we have in addition that 
\begin{equation}
W f(z_0)=\frac{\sigma}{\pi \hbar}~, \hspace{0.5 cm} \sigma=\pm 1~,
\label{eqrevision8A}
\end{equation}
then:

\begin{enumerate}
\item $\sigma =+1$, if $p=2^{k}$, for some $k\in \mathbb{N}$;

\item $\sigma =-1$, if $p$ is odd.
\end{enumerate}
\end{theorem}

\begin{remark}
\label{RemarkRationalRadius} It is fairly easy to construct functions that
satisfy the conditions of Theorem \ref{TheoremRationalRadius}. For instance,
the functions $f_{1}=\frac{1}{2}h_{2}+\frac{\sqrt{3}}{2}h_{4}$ and $f_{2}=%
\frac{1}{\sqrt{3}}h_{1}+\sqrt{\frac{2}{3}}h_{3}$ are such that: $Wf_{1}(0)=%
\frac{1}{\pi \hbar }$ and $Wf_{1}(z)$ vanishes on the circle $\frac{2|z|^{2}%
}{\hbar}=2$; while $Wf_{2}(0)=-\frac{1}{\pi \hbar }$ and $Wf_{2}(z)$
vanishes on the circle $\frac{2|z|^{2}}{\hbar}=3$.

On the other hand, there exists no Wigner function (with a bounded nodal
set), such that $Wf(0)=-\frac{1}{\pi \hbar }$ which vanishes on the circle $%
\frac{2|z|^{2}}{\hbar}=p$ with $p=2^{k}$, for some $k\in \mathbb{N}$, or a
Wigner function such that $Wf(0)=\frac{1}{\pi \hbar }$ which vanishes on the
circle $\frac{2|z|^{2}}{\hbar}=p$ with $p$ odd.
\end{remark}

By acting with time-frequency shifts and symplectic transformations, all the
previous results concerning circular zeros can be restated \textit{mutatis
mutandis} for elliptic zeros.

If a Wigner function $Wf$ has a bounded nodal set, then $f$ must be of the
form (\ref{eqTheoremEllipse1}). However, without further information there
is little that one can say regarding the `rank' $N$ in that expression. In
the next theorem, we show that `small' circular zeros lead to large rank $N$%
, but the same is also true for large radii.

\begin{theorem}
\label{TheoremHermite1} Let $Wf(z)$ be a Wigner function with a bounded
nodal set, and suppose that $\mathcal{N}(Wf)$ contains a circular zero of
radius $R>0$.

\begin{enumerate}
\item[(1)] We have the following estimate for the rank. 
\begin{equation}
N\geq \frac{R-2}{4}\text{.}  \label{eqTheoremRank1}
\end{equation}

\item[(2)] If, for some $k\in \mathbb{N}$, $R\leq \frac{1}{k}$, then 
\begin{equation}
N\geq k  \label{RankN_k}
\end{equation}
\end{enumerate}
\end{theorem}

\begin{remark}
\label{RemarkArcCircularZero} At this point it is worth making the following
important remark. In the various proofs where we assume the existence of a
circular zero, the main results are partly based on the fact that the
trigonometric polynomials $\left\{e^{i k \theta}~, ~k \in \mathbb{Z}
\right\} $ are linearly independent for $\theta \in \left[0, 2 \pi \right]$.
But in fact, they are linearly independent on any interval $\left[a, b %
\right]$, $a <b$. This entails that, when we assume that $\mathcal{N}(Wf)$
contains a circular zero, we could instead just assume that it contains an
arc of a circumference, of arbitrarily small length.
\end{remark}

\subsubsection{Geometry of the nodal set}

\label{SubSection1.2}

Our first result concerning the geometry of nodal sets is a corollary of
Theorem \ref{TheoremEllipse}. It implies that no Wigner function (with
bounded nodal set) can vanish on a circle and on an ellipse, or on two
ellipses whose axes are not parallel.

\begin{corollary}
\label{CorollaryEllipse} Let $Wf(z)$ be a Wigner function with a bounded
nodal set. Suppose that $\mathcal{N}(Wf)$ contains some ellipse $%
(z-z_{0})\cdot M(z-z_{0})=1$ (with $z_{0}\in \mathbb{R}^{2}$ and $M$ a $%
2\times 2$ real, symmetric, positive-definite matrix). Then $\mathcal{N}(Wf)$
can only contain other ellipses if they are of the form $\mu
(z-z_{0}^{\prime })\cdot M(z-z_{0}^{\prime })=1$, for some $z_{0}^{\prime
}\in \mathbb{R}^{2}$ and $\mu >0$.
\end{corollary}

The following theorem shows that Wigner functions with bounded nodal sets
cannot vanish on polygons or any curves which contain line segments\footnote{%
In \cite{Janssen1} the author proved that a Wigner distribution cannot be
concentrated on a curve, except if it is a straight line. See also \cite%
{Janssen3}.}.

\begin{theorem}
\label{TheoremLineSegment} Let $Wf(z)$ be a Wigner function with a bounded
nodal set. Then $\mathcal{N}(Wf)$ cannot contain any line segment.
\end{theorem}

Finally, Wigner functions can vanish at isolated points, but that poses some
restrictions on its rank:

\begin{proposition}
\label{PropositionIsolatedZeros} If $Wf$ has isolated zeros then it has rank 
$N \geq 3$.
\end{proposition}

\section{Notation and further remarks}

\label{Section2}

\subsection{Notation}

We denote by $x\cdot y=x_{1}y_{1}+\cdots x_{d}y_{d}$ the inner product of $%
x=(x_{1},\cdots ,x_{d}),y=(y_{1},\cdots ,y_{d})\in \mathbb{R}^{d}$. We use
the symbol $|x|=\sqrt{x_{1}^{2}+\cdots +x_{d}^{2}}$ for the norm of a vector 
$(x_{1},\cdots ,x_{d})\in \mathbb{R}^{d}$ and for the absolute value of a
complex number $|z|^{2}=z~\bar{z}$. The $L^{p}$ norm of a function is 
\begin{equation}
\Vert f\Vert _{p}=\left( \int_{\mathbb{R}^{d}}|f(x)|^{p}dx\right) ^{\frac{1}{%
p}}~,  \label{eqNotation1}
\end{equation}%
for $1\leq p<\infty $, and 
\begin{equation}
\Vert f\Vert _{\infty }=\text{ess sup}|f(x)|~,  \label{eqNotation2}
\end{equation}%
for $p=\infty $. The space of continuous complex-valued functions and the
Schwartz space of rapid-descreasing test functions on $\mathbb{R}^{d}$ will
be denoted by $C(\mathbb{R}^{d})$ and $\mathcal{S}(\mathbb{R}^{d})$,
respectively. We use latin upper-case letters A,B,C,... to enumerate known
theorems, and new results will be numerated as Theorem 1, Proposition 1, 
\textit{etc}. The set of positive integers is $\mathbb{N}=\left\{
1,2,3,\cdots \right\} $ and if we include zero, we write $\mathbb{N}%
_{0}=\left\{ 0,1,2,3,\cdots \right\} $.

\subsection{Wigner transforms of Hermite functions}

Wigner functions have positive marginals: 
\begin{equation}
\begin{array}{ll}
\int_{\mathbb{R}}Wf(x,p)dp=|f(x)|^{2} & ,~\forall x\in \mathbb{R} \\ 
&  \\ 
\int_{\mathbb{R}}Wf(x,p)dx=|\mathcal{F}_{\hbar }f(p)|^{2} & ,~\forall p\in 
\mathbb{R}%
\end{array}
\label{eqrevision4}
\end{equation}%
provided $f,\mathcal{F}_{\hbar }f\in L^{1}(\mathbb{R})\cap L^{2}(\mathbb{R})$%
. Here, $\mathcal{F}_{\hbar }f$ denotes the Fourier transform: 
\begin{equation}
(\mathcal{F}_{\hbar }f)(p)=\frac{1}{\sqrt{2\pi \hbar }}\int_{\mathbb{R}%
}f(x)e^{-\frac{i}{\hbar }xp}dx\text{~.}  \label{eqrevision5}
\end{equation}%
Wigner functions share many properties with ordinary probability densities.
They are real, and, as we assume in this paper, normalized to $1$: 
\begin{equation}
\int_{\mathbb{R}}\int_{\mathbb{R}}Wf(x,p)dxdp=\Vert f\Vert _{2}^{2}=1~.
\label{eqrevision3}
\end{equation}%
However, unlike probability densities, with the notable exception of the
Gaussian function, Wigner functions can assume negative values. A classical
example is provided by the Hermite functions: 
\begin{equation}
h_{n}(x)=\frac{1}{\sqrt[4]{\pi \hbar }\sqrt{2^{n}n!}}e^{-\frac{x^{2}}{2\hbar 
}}H_{n}\left( \frac{x}{\sqrt{\hbar }}\right) ~,n\in \mathbb{N}_{0}\text{,}
\label{eqWigner20}
\end{equation}%
where 
\begin{equation}
H_{n}(x)=(-1)^{n}e^{x^{2}}\frac{d^{n}}{dx^{n}}e^{-x^{2}}  \label{eqWigner21}
\end{equation}%
are the Hermite polynomials. The Hermite functions provide an orthonormal
basis for $L^{2}(\mathbb{R})$. Their Wigner distributions are given by: 
\begin{equation}
W(h_{n+k},h_{n})(z)=\frac{(-1)^{n}}{\pi \hbar }\sqrt{\frac{n!}{(n+k)!}}%
\left( \frac{2}{\hbar }\right) ^{k/2}\overline{z}^{k}L_{n}^{(k)}\left( \frac{%
2|z|^{2}}{\hbar }\right) e^{-\frac{|z|^{2}}{\hbar }},~n,k\in \mathbb{N}_{0}
\label{eqWigner22}
\end{equation}%
In particular for $k=0$: 
\begin{equation}
Wh_{n}(z)=\frac{(-1)^{n}}{\pi \hbar }L_{n}\left( \frac{2|z|^{2}}{\hbar }%
\right) e^{-\frac{|z|^{2}}{\hbar }}~,~n=0,1,2,\cdots  \label{eqWigner24}
\end{equation}%
The functions $W(h_{n},h_{n+k})$ are obtained from $\overline{W(f,g)}=W(g,f)$%
. Under our basic assumption of $Wf$ with a bounded nodal set, $f$ is of the
form (\ref{eqTheoremEllipse1}), and the above formulas show that its Wigner
function can be written as: 
\begin{equation}
Wf(z)=Wg_{N}(S(z-z_{1}))~,  \label{eqTheoremEllipse1A}
\end{equation}%
where%
\begin{equation*}
g_{N}(x)=\sum_{n=0}^{N}b_{n}h_{n}(x)~\text{.}
\end{equation*}%
It follows from (\ref{eqWigner22}) and (\ref{eqWigner24}) that $Wg_{N}$ can
be written explicitly as 
\begin{equation}
\begin{array}{c}
Wg_{N}(z)=\sum_{n,m=0}^{N}b_{n}\overline{b_{m}}W(h_{n},h_{m})(z) \\ 
\\ 
=\sum_{n=0}^{N}|b_{n}|^{2}\frac{(-1)^{n}}{\pi \hbar }e^{-\frac{|z|^{2}}{%
\hbar }}L_{n}\left( \frac{2|z|^{2}}{\hbar }\right) + \\ 
\\ 
+2Re\sum_{n=0}^{N-1}\sum_{k=1}^{N-n}b_{n}\overline{b_{n+k}}\frac{(-1)^{n}}{%
\pi \hbar }\sqrt{\frac{n!}{(n+k)!}}a^{k}e^{-\frac{|z|^{2}}{\hbar }%
}L_{n}^{(k)}\left( \frac{2|z|^{2}}{\hbar }\right) ~,%
\end{array}
\label{eqTheoremEllipse1B}
\end{equation}%
and 
\begin{equation}
a=\sqrt{\frac{2}{\hbar }}(x+ip)\text{.}  \label{eqWigner23}
\end{equation}%
In the previous expressions $L_{n}^{(k)}$ are the classical Laguerre
polynomials \cite{Gradshteyn,Szego}.

\subsection{Wigner transforms and polyentire functions}

\label{RemarkPolyanalytic}

Wigner functions are not analytic functions when regarded as functions of
the complex variable $a=\sqrt{\frac{2}{\hbar }}(x+ip)$. However, results
such as Theorem \ref{TheoremSmootDecay} and Corollary \ref%
{CorollaryUniquenessSet} retain some of the flavour of entire function
theory. This reflects their polyentire structure \cite{Abreu,Balk,Jaming1}
(in Balk's terminology, polyanalytic functions with $U=\mathbb{C}$ are
called polyentire functions). A function $F(z)=F(z,\overline{z})$, $z\in 
\mathbb{C}$, is polyentire of order $N\in \mathbb{N}$, if it satisfies the
higher order Cauchy-Riemann equation $\left( \partial _{\overline{z}}\right)
^{N}F(z,\overline{z})=0$ or, equivalently, if it can be written as $F(z,%
\overline{z})=\sum_{n=0}^{N-1}f_{n}(z)\bar{z}^{n}$, where $f_{n}(z)$ are
entire functions. For instance, $|z|^{2}$ is polyentire of order $2$. The
polyentire structure of Wigner functions with bounded nodal set follows from
their explicit expansion on phase space (\ref{eqTheoremEllipse1B}). Clearly $%
F(z)=e^{\frac{|z|^{2}}{\hbar}}Wf(z)$ is a polyentire function of order $N+1$
in the complex variable $z=x+ip $. Some information about the zeros can be
obtained for so-called reduced polyentire functions, namely those which can
be written in the form $F(z,\overline{z})=\sum%
\limits_{n=0}^{N-1}|z|^{2}f_{n}(z)$, for entire functions $f_{n}(z)$ but
this is not the case of the real function $F(z)$. Nevertheless, it is true
that if two polyentire functions coincide on a domain, then they are
identical everywhere \cite[Theorem 1.2]{Balk}, and, together with Theorem %
\ref{TheoremSmootDecay}, this provides an alternative proof of Corollary \ref%
{CorollaryUniquenessSet}. Polyentire functions can be understood as
restrictions of two variable entire functions $F_{M}(z,w)$ to the hyperplane 
$w=\overline{z}$ and several of their properties result from this
observation \cite[Chapter 4]{Balk}. The structure of the zeros of polyentire
functions is far from being understood. For instance, entire functions
vanishing in a set containing an accumulation point must vanish for all $%
z\in \mathbb{C}$. On the contrary, for polyanalytic functions, this is not
true. Indeed, the polyanalytic functions $z$ and $\bar{z}$ coincide on the
real axis, but are different elsewhere. Another property which can be
deduced from the polyentire polynomial structure of $e^{\frac{|z|^{2}}{\hbar}%
}Wf(z)$ is the following \cite[Chapter 4]{Balk}: if there exists a real
constant $A$ such that $\left\vert e^{\frac{|z|^{2}}{\hbar}}Wf(z)\right\vert
\leq A|z|^{n}$ outside a ball $B_{R_{0}}(0)$ for some $R_{0}>0$, then $N
\leq n$.

Functions of this type can be understood as a subclass of polyanalytic
functions, as, for instance, the structure of eigenfunctions of the Landau
operator with a magnetic field, known as true/pure polyanalytic functions 
\cite{Abreu,Jaming1}. Polyentire functions with a bounded nodal set must be
polynomials \cite[Chapter 4]{Balk} of a finite degree but, despite the
similarities of this statement with Theorem \ref{TheoremSmootDecay}, none of
the two results can be deduced from the other.

\section{Essential properties of the Wigner transform}

\label{Section3}

In this section we revise some known results and prove a few new ones which
will be useful for the proof of the results stated in the introduction.

Hudson's theorem \cite{Hudson,Janssen2}, which is stated here for reference,
asserts that the only functions whose Wigner distribution has no zeros - and
hence, in particular, no negative region - are generalized Gaussians.

\begin{lettertheorem}[Hudson]
\label{TheoremHudsonVersion2} Let $f\in L^{2}(\mathbb{R})$. Then $\mathcal{N}%
(Wf)=\emptyset $ if and only if $f$ is a generalized Gaussian: 
\begin{equation}
f(x)=e^{-\frac{a}{2\hbar }x^{2}+\frac{b}{\hbar }x+c}~,  \label{eqrevision8}
\end{equation}%
where $a,b,c,\in \mathbb{C}$ and $\text{Re}(a)>0$.
\end{lettertheorem}

\subsection{Global bounds for the Wigner transform}

We recall the following well-known theorem which is a straightforward
consequence of the Cauchy-Schwartz inequality \cite{Grossmann,Royer}:

\begin{letterprop}
\label{PropositionBoundWigner} The Wigner distribution $Wf$ of $f\in L^{2}(%
\mathbb{R})$ is an absolutely continuous and bounded function, such that 
\begin{equation}
\Vert Wf\Vert _{L^{\infty }}\leq \frac{1}{\pi \hbar }~.
\label{eqPropositionBoundWigner1}
\end{equation}%
Moreover, we have 
\begin{equation}
Wf(z_{0})=\pm \frac{1}{\pi \hbar }~,  \label{eqPropositionBoundWigner2}
\end{equation}%
for some $z_{0}=(x_{0},p_{0})\in \mathbb{R}^{2}$ if and only if 
\begin{equation}
f(x_{0}+x)e^{-\frac{i}{\hbar }(x_{0}+x)p_{0}}=\pm f(x_{0}-x)e^{-\frac{i}{%
\hbar }(x_{0}-x)p_{0}}~,  \label{eqPropositionBoundWigner3}
\end{equation}%
respectively. In other words, the function $f(x)e^{-\frac{i}{\hbar }xp_{0}}$
is even or odd with respect to the point $x_{0}$.
\end{letterprop}

\subsection{Covariance properties of the Wigner transform}

We shall be mainly concerned with the covariance properties of Wigner
distributions with respect to certain transformations. The fundamental
operators in time-frequency analysis are the translation 
\begin{equation}
(T_{x_{0}}f)(x)=f(x-x_{0})\text{,}  \label{eqWigner11}
\end{equation}%
and modulation operators 
\begin{equation}
(M_{p_{0}}f)(x)=e^{\frac{i}{\hbar }xp_{0}}f(x)\text{,}  \label{eqWigner2}
\end{equation}%
for some $x_{0},p_{0}\in \mathbb{R}$. These operators extend to unitary
operators in $L^{2}(\mathbb{R})$ and satisfy the commutation relations 
\begin{equation}
T_{x_{0}}M_{p_{0}}=e^{-\frac{i}{\hbar }x_{0}p_{0}}M_{p_{0}}T_{x_{0}}\text{.}
\label{eqWigner3}
\end{equation}%
If we write $z=(x_{0},p_{0})$ we can denote their collective action as: 
\begin{equation}
\left( \pi (z_{0})f\right) (x):=\left( M_{p_{0}}T_{x_{0}}f\right) (x)=e^{%
\frac{i}{\hbar }p_{0}x}f(x-x_{0})\text{.}  \label{eqWigner4}
\end{equation}%
The composition of two time-frequency shifts is given by: 
\begin{equation}
\pi (z_{1})\pi (z_{2})=e^{-\frac{i}{\hbar }x_{1}p_{2}}\pi (z_{1}+z_{2})\text{%
,}  \label{eqWigner4A}
\end{equation}%
for $z_{1}=(x_{1},p_{1})$ and $z_{1}=(x_{2},p_{2})$. From this identity we
also conclude that: 
\begin{equation}
\left( \pi (z_{1})\right) ^{-1}=e^{-\frac{i}{\hbar }x_{1}p_{1}}\pi (-z_{1})%
\text{.}  \label{eqWigner4B}
\end{equation}%
A straightforward computation shows how the Wigner distribution behaves
under these time-frequency shifts: 
\begin{equation}
\begin{array}{c}
W\left( \pi (z_{1})f,\pi (z_{2})g\right) (x,p)= \\ 
\\ 
=\exp \left\{ \frac{i}{\hbar }\left[ x(p_{1}-p_{2})-\left( p-\frac{%
p_{1}+p_{2}}{2}\right) (x_{1}-x_{2})\right] \right\} \times \\ 
\\ 
\times W(f,g)\left( x-\frac{x_{1}+x_{2}}{2},p-\frac{p_{1}+p_{2}}{2}\right) 
\text{,}%
\end{array}
\label{eqWigner5}
\end{equation}%
for $z_{1}=(x_{1},p_{1})$ and $z_{2}=(x_{2},p_{2})$. Thus, in particular 
\begin{equation}
W(\pi (z_{0})f)(x,p)=Wf(z-z_{0})\text{.}  \label{eqWigner6}
\end{equation}%
Let $S\in Sp(2;\mathbb{R})$ be a symplectic matrix, that is a real $2\times
2 $ matrix, such that 
\begin{equation}
SJS^{T}=J\text{,}  \label{eqWigner7}
\end{equation}%
where 
\begin{equation}
J=\left( 
\begin{array}{cc}
0 & 1 \\ 
-1 & 0%
\end{array}%
\right)  \label{eqWigner8}
\end{equation}%
is the standard symplectic matrix. The standard symplectic form is given by: 
\begin{equation}
\sigma (z,z^{\prime })=z^{\prime }\cdot Jz=px^{\prime }-xp^{\prime }\text{,}
\label{eqWigner8A}
\end{equation}%
The Shale-Weyl representation $\mu (S)$ (which is determined up to an
overall sign) satisfies \cite{Gosson,Shale,Weil}: 
\begin{equation}
\mu (S)\pi (z_{0})\left( \mu (S)\right) ^{-1}=\pi (Sz_{0})\text{.}
\label{eqWigner9}
\end{equation}%
The Wigner function transforms as: 
\begin{equation}
W\left( \mu (S)f,\mu (S)g\right) (z)=W(f,g)(S^{-1}z)\text{,}
\label{eqWigner10}
\end{equation}%
for all $z=(x,p)\in \mathbb{R}^{2}$. We shall also consider the reflection
operator 
\begin{equation}
\left( \mathcal{I}f\right) (x)=f(-x)\text{,}  \label{eqWigner10A}
\end{equation}%
the dilation operator 
\begin{equation}
\left( D_{s}f\right) (x)=\sqrt{|s|}f(sx)\text{,}  \label{eqWigner11}
\end{equation}%
for $s\neq 0$, and the multiplication by a chirp: 
\begin{equation}
\left( \mathcal{N}_{\xi }f\right) (x)=e^{-\frac{i}{2\hbar }\xi x^{2}}f(x)%
\text{,}  \label{eqWigner12}
\end{equation}%
for $\xi \in \mathbb{R}$. Under these operations the Wigner distribution
behaves as follows (see Lemma 4.4.3 in \cite{Grochenig}): 
\begin{equation}
\begin{array}{l}
W(\mathcal{I}f,\mathcal{I}g)(x,p)=W(f,g)(-x,-p)\text{,} \\ 
\\ 
W(D_{s}f,D_{s}g)(x,p)=W(f,g)\left( sx,\frac{p}{s}\right) \text{,} \\ 
\\ 
W\left( \mathcal{N}_{\xi }f,\mathcal{N}_{\xi }g\right) (x,p)=W(f,g)(x,p-\xi
x)\text{.}%
\end{array}
\label{eqWigner13}
\end{equation}%
For future reference we recall that \cite{Folland}: 
\begin{equation}
\mathcal{I}\pi (z_{0})=\pi (-z_{0})\mathcal{I}~,~\text{ and }\mathcal{I}\mu
(S)=\mu (S)\mathcal{I}\text{.}  \label{eqWigner13A}
\end{equation}%
Notice that the generalized Gaussian (\ref{eqrevision8}) can be written as
(see Lemma 4.4.2 in \cite{Grochenig}): 
\begin{equation}
f(x)=k\left( M_{b_{2}-\frac{b_{1}}{s}\xi }T_{\frac{b_{1}}{s}}\mathcal{N}%
_{\xi }D_{\sqrt{s}}h_{0}\right) (x)\text{,}  \label{eqWigner14}
\end{equation}%
where $a=s+i\xi $, with $s>0$, $b=b_{1}+ib_{2}$, $b_{1},b_{2}\in \mathbb{C}$%
, $k\in \mathbb{C}$ and 
\begin{equation}
h_{0}(x)=\frac{1}{\sqrt[4]{\pi \hbar }}e^{-\frac{x^{2}}{2\hbar }}
\label{eqWigner15}
\end{equation}%
is the Gaussian Hermite function. It follows from (\ref{eqWigner6}, \ref%
{eqWigner13}) that, for $f$ of the form (\ref{eqWigner14}): 
\begin{equation}
\begin{array}{c}
Wf(x,p)=|k|^{2}Wh_{0}\left( \sqrt{s}x-\frac{b_{1}}{\sqrt{s}},\frac{p}{\sqrt{s%
}}-\frac{\xi x}{\sqrt{s}}-\frac{b_{2}}{\sqrt{s}} +\frac{2b_1 \xi}{s \sqrt{s}}%
\right) \\ 
\\ 
\Leftrightarrow Wf(z)=Wh_{0}\left( Sz-z_{0}\right) \text{,}%
\end{array}
\label{eqWigner16}
\end{equation}%
where we normalized the function, and 
\begin{equation}
z_{0}=\left( \frac{b_{1}}{\sqrt{s}},\frac{b_{2}-2b_1 \xi /s}{\sqrt{s}}%
\right) \text{,}  \label{eqWigner16A}
\end{equation}%
and $S$ is the symplectic matrix 
\begin{equation}
S=\left( 
\begin{array}{cc}
\sqrt{s} & 0 \\ 
-\frac{\xi }{\sqrt{s}} & \frac{1}{\sqrt{s}}%
\end{array}%
\right) \text{.}  \label{eqWigner16B}
\end{equation}%
Finally, since 
\begin{equation}
Wh_{0}(z)=\frac{1}{\pi \hbar }e^{-\frac{|z|^{2}}{\hbar }}\text{,}
\label{eqWigner17}
\end{equation}%
we obtain: 
\begin{equation}
Wf(z)=Wh_{0}\left( Sz-z_{0}\right) =\frac{1}{\pi \hbar }e^{-\frac{1}{\hbar }%
|Sz-z_{0}|^{2}}=\frac{1}{\pi \hbar }e^{-\frac{1}{\hbar }(z-z_{1})\cdot
M(z-z_{1})}\text{,}  \label{eqWigner18}
\end{equation}%
where $z_{1}=S^{-1}z_{0}$ and 
\begin{equation}
M=S^{T}S\in Sp(2;\mathbb{R})\text{.}  \label{eqWigner19}
\end{equation}

\subsection{Husimi function and Bargmann transform}

The Husimi function \cite{Husimi} of $f\in L^{2}(\mathbb{R})$ is given by
the convolution of the Wigner function $Wf$ with $Wh_{0}$: 
\begin{equation}
Hf(z):=\left( Wf\star Wh_{0}\right) (z)=\frac{1}{\pi \hbar }\int_{\mathbb{R}%
^{2}}Wf(\omega )e^{-\frac{1}{\hbar }|z-\omega |^{2}}d\omega \text{.}
\label{eqHusimi1}
\end{equation}%
The Husimi function (the spectrogram with window $h_{0}$) can be expressed
as 
\begin{equation}
Hf(x,p)=\left\vert \mathcal{B}f\left( \frac{x-ip}{\sqrt{2}}\right)
\right\vert ^{2}\frac{e^{-\frac{|z|^{2}}{2\hbar }}}{2\pi \hbar }\text{,}
\label{eqHusimi2}
\end{equation}%
in terms of the Bargmann-Segal transform \cite{Bargmann,Folland,Hall,Segal}: 
\begin{equation}
\mathcal{B}f(\zeta ):=\frac{1}{\sqrt[4]{\pi \hbar }}\int_{\mathbb{R}}e^{-%
\frac{1}{2\hbar }\zeta ^{2}-\frac{\sqrt{2}}{\hbar }x\zeta -\frac{x^{2}}{%
2\hbar }}f(x)dx\text{,}  \label{eqHusimi3}
\end{equation}%
for $\zeta \in \mathbb{C}$. The Bargmann transform is an entire function
with order of growth at most $2$.

\subsection{Pseudo-displacement operator}

\label{subsectionPseudoDispOp}

\label{SubsectionPseudoDispOp}

In this section we discuss the properties of an operator, which some of us
introduced recently in \cite{Zacharie}. We called it \textit{%
pseudo-displacement} operator, because when acting on a Wigner function
which is the product of a polynomial function and a Gaussian, it produces a
displacement (time-frequency shift) of the polynomial part, while leaving
the Gaussian part invariant.

We start by considering the following multiplication operator, for some $%
\alpha \in \mathbb{C}$ 
\begin{equation}
\left(\mathcal{M}_{\alpha} f \right)(x)= e^{\alpha x} f(x)~.
\label{eqpseudoDisp1}
\end{equation}
If $\alpha =\frac{i}{\hbar}p_0$ is purely imaginary, then we just have a
modulation. Notice that for $\text{Re}(\alpha) \neq 0$ this not closed in $%
L^2 (\mathbb{R})$. For example, if $f(x)= e^{- \beta |x|} $ is such that $0
< \text{Re} (\beta) \leq \text{Re} (\alpha)$, then $f \in L^2 (\mathbb{R})$,
but $\mathcal{M}_{\alpha} f \notin L^2 (\mathbb{R})$. However, for the
purposes of the present paper it is sufficient to work with the following
finite dimensional Hilbert spaces.

Let $z_1=(x_1,p_1) \in \mathbb{R}^2$ and $N \in \mathbb{N}$. Let us define
the Hilbert space: 
\begin{equation}
\mathcal{H}_{z_1}^N:= \text{span} \left\{\pi (z_1) h_k:~ 0 \leq k \leq N
\right\}~.  \label{eqpseudoDisp2}
\end{equation}
We recall that $h_k(x) = \frac{1}{\sqrt[4]{\pi \hbar} \sqrt{2^k k!}} H_k
\left(\frac{x}{\sqrt{\hbar}} \right) e^{- \frac{x^2}{2 \hbar}}$ and that $%
\text{span}\left\{H_k:~ 0\leq k\leq N \right\}= \mathbb{C}_N\left[x\right]$,
the vector space of polynomials of degree at most $N$ with complex
coefficients. Hence, $\mathcal{H}^N_{z_1}$ can also be written in the form: 
\begin{equation}  \label{newH}
\mathcal{H}^N_{z_1}= \left\{P_N(x) e^{-\frac{(x-x_1)^2}{2 \hbar} } e^{\frac{i%
}{\hbar}xp_1}:~ P_N(x) \in \mathbb{C}_N\left[x\right] \right\}
\end{equation}
where we used the fact that, if $Q_N(x) \in \mathbb{C}_N\left[x\right]$,
then there exists $P_N(x) \in \mathbb{C}_N\left[x\right]$ such that $%
Q_N(x-x_1)=P_N(x)$.

We then have:

\begin{proposition}
\label{PropositionpseudoDisp1} The following maps are isomorphisms. 
\begin{equation}
\begin{array}{ll}
(i) & \mathcal{M}_{\alpha}: \mathcal{H}_{z_1}^N \to \mathcal{H}_{z_2}^N~, ~%
\text{ where } z_2=\left(x_1+ \hbar \text{Re} (\alpha),p_1 + \hbar \text{Im}%
(\alpha) \right) \\ 
&  \\ 
(ii) & \mathcal{F}_{\hbar}: \mathcal{H}_{z_1}^N \to \mathcal{H}_{J z_1}^N \\ 
&  \\ 
(iii) & \mathcal{I}: \mathcal{H}_{z_1}^N \to \mathcal{H}_{-z_1}^N~%
\end{array}
\label{eqpseudoDisp3}
\end{equation}
\end{proposition}

\begin{proof}
If $f \in \mathcal{H}_{z_1}^N$, then
\begin{equation}
f= \pi(z_1) \sum_{n=0}^N c_n h_n ~, 
\label{eqpseudoDisp4}
\end{equation}
for some $\left\{c_j:~j=0, \cdots, N \right\} \subset \mathbb{C}$.

\begin{enumerate}
\item [(i)] Let us set $\alpha= \alpha_1 + i \alpha_2$. We have:
\begin{equation}
\begin{array}{c}
\left(\mathcal{M}_{\alpha}f\right)(x)=e^{\alpha_1 x} \sum_{n=0}^N c_n e^{\frac{i}{\hbar}x(p_1+ \hbar \alpha_2)} h_n (x-x_1)=\\
\\
= \sum_{n=0}^N \frac{c_n}{\sqrt[4]{\pi \hbar} \sqrt{2^n n !}}  H_n \left(\frac{x-x_1}{\sqrt{\hbar}} \right)  e^{- \frac{(x-x_1)^2}{2 \hbar} + \alpha_1 x} e^{\frac{i}{\hbar}x(p_1+ \hbar \alpha_2)} 
\end{array}
\label{eqpseudoDisp5}
\end{equation}
Setting $z_2=(x_1+ \hbar \alpha_1,p_1+ \hbar \alpha_2)$ and using the fact that 
\begin{equation}
- \frac{1}{2 \hbar}(x-x_1)^2 + \alpha_1 x=- \frac{1}{2 \hbar}(x-x_1- \hbar \alpha_1)^2 + \frac{\hbar}{2}\alpha_1^2 +  \alpha_1 x_1~,  
\label{eqpseudoDisp7}
\end{equation}
we conclude from (\ref{newH}) that:
\begin{equation}
\left(\mathcal{M}_{\alpha}f\right)(x)
= \left( \pi (z_2) \sum_{k=0}^N b_k h_k \right)(x)~,
\label{eqpseudoDisp8}
\end{equation}
for some complex constants $(b_k)$.

Thus, $\mathcal{M}_{\alpha}f \in \mathcal{H}_{z_2}^N$. By replacing $z_1$ by $z_2$ and $\alpha$ by $- \alpha$ it is obvious that $\mathcal{M}_{\alpha}$ has inverse $\mathcal{M}_{-\alpha}$ and hence it is an isomorphism.

\vspace{0.2 cm}
\item [(ii)] We use the commutation relations
\begin{equation}
\mathcal{F}_{\hbar} M_{p_1}=T_{p_1}\mathcal{F}_{\hbar}~, \hspace{0.5 cm}\mathcal{F}_{\hbar} T_{x_1}=M_{- x_1}\mathcal{F}_{\hbar}~,
\label{eqpseudoDisp9}
\end{equation}
and the fact that Hermite functions are eigenfunctions of the Fourier transform
\begin{equation}
\mathcal{F}_{\hbar}  h_n=(-i)^n h_n ~.
\label{eqpseudoDisp10}
\end{equation}
We thus have:
\begin{equation}
\begin{array}{c}
\mathcal{F}_{\hbar} f= \mathcal{F}_{\hbar} M_{p_1}T_{x_1} \sum_{n=0}^N  c_n h_n=\\
\\
= T_{p_1} \mathcal{F}_{\hbar} T_{x_1} \sum_{n=0}^N  c_n h_n=\\
\\
=T_{p_1}  M_{-x_1} \sum_{n=0}^N  c_n   \mathcal{F}_{\hbar}h_n=\\
\\
= e^{\frac{i}{\hbar} p_1x_1} M_{-x_1} T_{p_1}   \sum_{n=0}^N  c_n   \mathcal{F}_{\hbar}h_n=\\
\\
=\pi(Jz_1) \sum_{n=0}^N  c_n   (-i)^n e^{\frac{i}{\hbar} p_1x_1} h_n ~. 
\end{array}
\label{eqpseudoDisp11}
\end{equation}
We conclude that $\mathcal{F}_{\hbar} f\in \mathcal{H}_{J z_1}^N$. That this is an isomorphism is a simple consequence of the fact that the Fourier transform is invertible in $\mathcal{S}(\mathbb{R})$ and that it is surjective onto $\mathcal{H}_{Jz_1}^N$.

\vspace{0.2 cm}
\item [(iii)] We use the commutation relations
\begin{equation}
\mathcal{I} \pi(z_1)=\pi(-z_1)\mathcal{I}~,
\label{eqpseudoDisp12}
\end{equation}
and the fact that $h_n$ are even or odd functions:
\begin{equation}
\mathcal{I}h_n=(-1)^n h_n~.
\label{eqpseudoDisp13}
\end{equation}
Next:
\begin{equation}
\begin{array}{c}
\mathcal{I} f= \mathcal{I} \pi(z_1) \sum_{n=0}^N  c_n h_n=\\
\\
=\pi(-z_1) \sum_{n=0}^N  c_n   \mathcal{I} h_n=\pi(-z_1) \sum_{n=0}^N  c_n (-1)^n h_n~.
\end{array}
\label{eqpseudoDisp14}
\end{equation}
Thus, $\mathcal{I} f \in \mathcal{H}_{- z_1}^N$. Since $\mathcal{I}$ is invertible in $L^2(\mathbb{R})$ and surjective onto $\mathcal{H}_{-z_1}^N$, we have an isomomorphism. 
\end{enumerate}

\end{proof}

We will henceforth assume that $f \in \mathcal{H}_{z_1}^N$ for some $z_1$
and some $N$.

Next, we consider the effect of this multiplication operator under the
Wigner transform. A straightforward calculation shows that 
\begin{equation}
W\left(\mathcal{M}_{\alpha} f \right)(x,p)= e^{2 \text{Re} (\alpha) x}~
Wf(x,p- \hbar \text{Im} (\alpha))~.  \label{eqpseudoDisp15}
\end{equation}
Let us now define the \textit{pseudo-displacement operator} 
\begin{equation}
\left(D_{\alpha} f \right)(x):= \left(\mathcal{M}_{\alpha}\mathcal{F}%
_{\hbar}^{-1} \mathcal{M}_{i \alpha} \mathcal{F}_{\hbar} f \right)(x)~.
\label{eqpseudoDisp16}
\end{equation}

\begin{lemma}
\label{LemmaPseudoDisp1} The map 
\begin{equation}
D_{\alpha}: \mathcal{H}_{z_1}^N \to \mathcal{H}_{z_1}^N~
\label{eqpseudoDisp17}
\end{equation}
is a bijection.
\end{lemma}

\begin{proof}
This is a simple consequence of Proposition \ref{PropositionpseudoDisp1} and the fact that $\mathcal{F}_{\hbar}^{-1}= \mathcal{I}\mathcal{F}_{\hbar}$.
\end{proof}

\begin{proposition}
\label{PropostionPseudoDisp3} Let $f \in \mathcal{H}_{z_1}^N$ and let $%
\alpha=\frac{1}{\hbar}(-x_0+i p_0)$. Set $z=(x,p)$ and $z_0=(x_0,p_0)$.
Then: 
\begin{equation}
W\left(D_{\alpha} f \right)(z)= e^{\frac{2}{\hbar} (p_0^2-z_0 \cdot
z)}~Wf(z-z_0)~.  \label{eqpseudoDisp18}
\end{equation}
\end{proposition}

\begin{proof}
As previously, we set $\alpha_1= \text{Re} (\alpha)$ and $\alpha_2= \text{Im} (\alpha)$. Taking into account the fact that $W\left(\mathcal{F}_{\hbar} f\right)(z)=Wf(- Jz)$, we obtain:
\begin{equation}
\begin{array}{c}
W\left(D_{\alpha} f \right)(x,p)= W \left(\mathcal{M}_{\alpha}\mathcal{F}_{\hbar}^{-1} \mathcal{M}_{i \alpha} \mathcal{F}_{\hbar} f \right)(x,p)=\\
\\
= e^{2 \alpha_1 x}~W \left(\mathcal{F}_{\hbar}^{-1} \mathcal{M}_{i \alpha} \mathcal{F}_{\hbar} f \right)(x,p- \hbar \alpha_2)=\\
\\
= e^{2 \alpha_1 x}~W \left(\mathcal{F}_{\hbar} \mathcal{M}_{i \alpha} \mathcal{F}_{\hbar} f \right)(-x,\hbar \alpha_2-p)=\\
\\
= e^{2 \alpha_1 x}~W \left( \mathcal{M}_{i \alpha} \mathcal{F}_{\hbar} f \right)(p-\hbar \alpha_2,-x)=\\
\\
= e^{2 \alpha_1 x} e^{-2 \alpha_2 (p-\hbar \alpha_2)} ~W \left( \mathcal{F}_{\hbar} f \right)(p-\hbar \alpha_2,-x - \hbar \alpha_1)=\\
\\
=e^{2 \alpha_1 x-2 \alpha_2 p} e^{2 \hbar \alpha_2^2} ~W f(x + \hbar \alpha_1,p- \hbar \alpha_2)~.
\end{array}
\label{eqpseudoDisp19}
\end{equation}
If we replace $\alpha_1= -\frac{x_0}{\hbar}$ and $\alpha_2= \frac{p_0}{\hbar}$, the result follows.
\end{proof}

Let us briefly comment on the action of the pseudo-displacement operator on
the Wigner function. Suppose that $f\in \mathcal{H}_{z_{1}}^{N}$. Then $Wf$
is of the form 
\begin{equation}
Wf(z)=P_{N}(z-z_{1})e^{-\frac{(z-z_{1})^{2}}{\hbar }}~,
\label{eqpseudoDisp20}
\end{equation}%
where $P_{N}(\cdot )$ is a polynomial of degree $N$. Under the action of $%
D_{\alpha }$, we have (cf.(\ref{eqpseudoDisp18})): 
\begin{equation}
W\left( D_{\alpha }f\right) (z)=e^{-\frac{1}{\hbar }(z_{0}^{2}+2z_{0}\cdot
z_{1}-2p_{0}^{2})}\left[ P_{N}(z-z_{1}-z_{0})e^{-\frac{1}{\hbar }%
(z-z_{1})^{2}}\right] ~.  \label{eqpseudoDisp21}
\end{equation}%
Thus, up to an unimportant multiplicative constant, the action of $D_{\alpha
}$ on $Wf$ amounts to a translation of the polynomial part, while leaving
the Gaussian part unchanged.

Let us also introduce the map:

\begin{definition}
\label{DefinitionOperatorE} For $z_0=(x_0,p_0) \in \mathbb{R}^2$, let: 
\begin{equation*}
E_{z_0}: \mathcal{H}^N_{z_1} \to \mathcal{H}^N_{z_1+z_0}~ , \hspace{0.3 cm}
E_{z_0}:= \pi(z_0) D_{-\alpha}
\end{equation*}
where $\alpha=\tfrac{1}{\hbar}(-x_0+ip_0)$.
\end{definition}

We leave to the reader the simple proofs of the following theorem and its
two corollaries.

\begin{theorem}

We have:

\begin{enumerate}
\item $E_{z_0}$ is a bijection.

\item If $f \in \mathcal{H}_{z_1}^N$ then $Wf(z)=P_N(z-z_1)e^{-\frac{%
(z-z_1)^2}{\hbar}}$ and 
\begin{equation*}
W\left(E_{z_0} f\right)(z)=K P_N(z-z_1)e^{- \frac{(z-z_1-z_0)^2}{\hbar}}
\end{equation*}
for some constant $K >0$. Here $P_N(z)$ is real polynomial of degree $N$.

\item Let $g=E_{z_0}f$. Then ${\mathcal{N}}(Wf)={\mathcal{N}}(Wg)$.
\end{enumerate}
\end{theorem}

It follows that:

\begin{corollary}
$f\in \mathcal{H}_{z_1}^N$ if and only if $g:=E_{-z_1}f \in \mathcal{H}_0^N$%
. Moreover: 
\begin{equation*}
{\mathcal{N}}(Wf)={\mathcal{N}}(Wg).
\end{equation*}
\end{corollary}

We then conclude the following.

\begin{corollary}
\label{CorollarypseudoDisp1} Suppose that $f \in \mathcal{H}_{z_1}^N$. Then,
there exists $g \in \mathcal{H}_{0}^N$, and a constant $K \in \mathbb{R}^+$,
such that $\mathcal{N}(Wf)= \mathcal{N}(Wg)$, and 
\begin{equation}
Wf(z)=K Wg(z) e^{- \frac{1}{\hbar}(z_1^2 -2 z_1 z)}~.  \label{eqpseudoDisp22}
\end{equation}
\end{corollary}

\section{Laguerre polynomials}

\label{SectionLaguerrePolynomials}

For $n=0,1,2,\cdots $ and $\alpha >-1$, the generalized Laguerre polynomials 
$L_{n}^{(\alpha )}(x)$ can be defined as: 
\begin{equation}
L_{n}^{(\alpha )}(x)=\sum_{j=0}^{n}(-1)^{j}\left( 
\begin{array}{c}
n+\alpha \\ 
n-j%
\end{array}%
\right) \frac{x^{j}}{j!}~,  \label{eqLaguerre1}
\end{equation}%
where as usual 
\begin{equation}
\left( 
\begin{array}{c}
\alpha \\ 
n%
\end{array}%
\right) =\frac{\alpha (\alpha -1)(\alpha -2)\cdots (\alpha -n+1)}{%
n(n-1)(n-2)\cdots 2\cdot 1}~,  \label{eqLaguerre2}
\end{equation}%
is a generalized binomial coefficient for $n\in \mathbb{N}$ and arbitrary $%
\alpha \in \mathbb{R}$. It is well-known that, for $\alpha >-1$, the zeros
of the Laguerre polynomials $L_{n}^{(\alpha )}$ are exclusively real, simple
and positive and, from a result (Theorem 3.3.2 \cite{Tricomi}) in the
general theory of orthogonal polynomials, that the zeros of $L_{n}^{(\alpha
)}$ interlace with those of $L_{n-1}^{(\alpha )}$, in the sense that between
any two successive zeros of $L_{n}^{(\alpha )}$ there is exactly one zero of 
$L_{n-1}^{(\alpha )}$, and conversely \cite{Driver1}. The Laguerre
polynomials of order zero, $L_{n}:=L_{n}^{(0)}$, reads: 
\begin{equation}
L_{n}(x)=\sum_{j=0}^{n}(-1)^{j}\left( 
\begin{array}{c}
n \\ 
j%
\end{array}%
\right) \frac{x^{j}}{j!}~.  \label{eqLaguerre3}
\end{equation}%
We shall denote the zeros of $L_{n}$, with $n\geq 1$, in increasing order
by: 
\begin{equation}
0<\rho _{1}^{(n)}<\rho _{2}^{(n)}<\cdots <\rho _{n}^{(n)}~.
\label{eqLaguerre4}
\end{equation}%
More generally, we shall denote by $\rho_{j, \alpha}^{(n)}$ the j-th root of 
$L_n^{(\alpha)}$.

The generalized Laguerre polynomial $Y=L_N^{(\alpha)}$, with $\alpha >-1$
satisfies the differential equation: 
\begin{equation}
xY^{\prime \prime}+(\alpha+1-x) Y^{\prime} +NY=0~.  \label{eqLaguerreDiffEq}
\end{equation}

The following result will be useful in the proof of Theorem \ref%
{TheoremRationalRadius}.

\begin{proposition}
\label{PropositionLaguerrezeros3} Let $n\in \mathbb{N}$, $m \in \mathbb{N}_0$
and suppose that $L_n^{(m)}(x)$ has a rational root $\bar x$. Then, we must
have $\bar x \in \mathbb{N}$. Moreover if $\bar x \in \mathbb{N}$ and we
write $\bar x= p_1^{n_1} \cdots p_k^{n_k}$ as the prime factor decomposition
of $\bar x$, then $L_n^{(m)}(\bar x)=0$ is possible only if $n=p_1^{l_1}
\cdots p_k^{l_k}$, for some $l_1, \cdots, l_k \in \mathbb{N}_0$.
\end{proposition}

\section{Proofs of main results}

\label{Section4}

\subsection{Proof of Theorem \protect\ref{TheoremConvolution}}

\label{SubsectionConvolution}

The convolution of two functions is defined as: 
\begin{equation}
(f\star g)(x):=\int_{\mathbb{R}^{n}}f(x-y)g(y)dy=\int_{\mathbb{R}%
^{n}}f(y)g(x-y)dy~.  \label{eqDefconv1}
\end{equation}%
Young's convolution inequality states that\footnote{%
For optimal constants see e.g. \cite{Babenko,Barthe,Beckner,Lieb1}.} 
\begin{equation}
\Vert f\star g\Vert _{r}\leq \Vert f\Vert _{p}\Vert g\Vert _{q}~,
\label{eqDefconv2}
\end{equation}%
for all $r,p,q\in \left[ 1,\infty \right] $ such that $\frac{1}{p}+\frac{1}{q%
}=1+\frac{1}{r}$. Thus, it is a continuous map $L^{1}(\mathbb{R}^{n})\times
L^{2}(\mathbb{R}^{n})\rightarrow L^{2}(\mathbb{R}^{n})$ and $L^{1}(\mathbb{R}%
^{n})\times L^{\infty }(\mathbb{R}^{n})\rightarrow L^{\infty }(\mathbb{R}%
^{n})$.

\textbf{Step 1. }We will use the notation $\Sigma (x)=(f\star g)(x)$. If $%
f\in L^{\infty }(\mathbb{R}^{n})$, then $\Sigma (x)$ is well defined for
every $x\in \mathbb{R}^{n}$. Since the nodal set of $f$ is bounded, there
exists $R_{0}>0$ such that: 
\begin{equation}
\mathcal{N}(f)\subset B_{R_{0}}=\left\{ x\in \mathbb{R}^{n}:|x|\leq
R_{0}\right\} ~.  \label{eqTheoremConvolution1}
\end{equation}%
Let us start with the case $n=1$. This case has to be treated separately,
because in one dimension a continuous function may have opposite signs for $%
x<-R_{0}$ and $x>R_{0}$, whereas in dimension $n>1$, the sign outside of the
ball $B_{R_{0}}$ is constant. We will prove that there exists $R>R_{0}$,
such that $\Sigma (x)\neq 0$ for all $x>R$. For $x<-R$ this is proved
analogously. We shall assume that $f(x)>0$ for all $x>R_{0}$. If this is not
the case, then we simply consider the function $-f(x)$, instead. Fix some $%
\epsilon >0$ and assume that $x>R_{0}+3\epsilon $. We then have: 
\begin{equation}
\begin{array}{c}
\left\vert \int_{-\infty }^{R_{0}}f(y)e^{-(x-y)^{2}}dy\right\vert \leq \Vert
f\Vert _{\infty }\int_{-\infty }^{R_{0}}e^{-(x-y)^{2}}dy= \\ 
\\ 
=\Vert f\Vert _{\infty }\int_{-\infty }^{R_{0}+\epsilon }e^{-(x-\tau
+\epsilon )^{2}}d\tau \leq \Vert f\Vert _{\infty }e^{-(x+\epsilon
)^{2}+2(R_{0}+\epsilon )(x+\epsilon )}\int_{-\infty }^{R_{0}+\epsilon
}e^{-\tau ^{2}}d\tau \\ 
\\ 
\leq \sqrt{\pi }\Vert f\Vert _{\infty }e^{-x^{2}+2R_{0}x+\epsilon
^{2}+2R_{0}\epsilon }~.%
\end{array}
\label{eqTheoremConvolution2}
\end{equation}

On the other hand: 
\begin{equation}
\begin{array}{c}
\int_{R_{0}}^{+\infty }f(y)e^{-(x-y)^{2}}dy\geq \int_{R_{0}+\epsilon
}^{R_{0}+3\epsilon }f(y)e^{-(x-y)^{2}}dy \\ 
\geq m_{\epsilon }e^{-(x-R_{0}-\epsilon )^{2}}=\frac{m_{\epsilon
}e^{-x^{2}+2x(R_{0}+\epsilon )}}{e^{(R_{0}+\epsilon )^{2}}}%
\end{array}
\label{eqTheoremConvolution3}
\end{equation}%
where $m_{\epsilon }=2\epsilon ~\text{min}\left\{ f(y):~R_{0}+\epsilon \leq
y\leq R_{0}+3\epsilon \right\} $. Thus: 
\begin{equation}
\begin{array}{c}
\Sigma (x)=\int_{-\infty }^{R_{0}}f(y)e^{-(x-y)^{2}}dy+\int_{R_{0}}^{+\infty
}f(y)e^{-(x-y)^{2}}dy \\ 
\\ 
\geq \frac{m_{\epsilon }e^{-x^{2}+2x(R_{0}+\epsilon )}}{e^{(R_{0}+\epsilon
)^{2}}}-\sqrt{\pi }\Vert f\Vert _{\infty }e^{-x^{2}+2R_{0}x+\epsilon
^{2}+2R_{0}\epsilon } \\ 
=e^{-x^{2}+2R_{0}x}\left[ \frac{m_{\epsilon }e^{2\epsilon x}}{%
e^{(R_{0}+\epsilon )^{2}}}-\sqrt{\pi }\Vert f\Vert _{\infty }e^{\epsilon
^{2}+2R_{0}\epsilon }\right]%
\end{array}
\label{eqTheoremConvolution4}
\end{equation}%
Since $m_{\epsilon }>0$, it is clear that $\Sigma (x)>0$, for all $x>R$,
where 
\begin{equation}
R=\text{max}\left\{ R_{0}+3\epsilon ,\frac{1}{2\epsilon }\left[ \ln \left( 
\frac{\sqrt{\pi }\Vert f\Vert _{\infty }}{m_{\epsilon }}\right) +2\epsilon
^{2}+4R_{0}\epsilon +R_{0}^{2}\right] \right\} ~.
\label{eqTheoremConvolution5}
\end{equation}

\textbf{Step 2. }Next we consider the case $n>1$. Without loss of
generality, we assume again that $f(x)>0$ for all $|x|>R_{0}$. Set $r=|x|$
to yield: 
\begin{equation}
\begin{array}{c}
\left\vert \int_{B_{R_{0}}}e^{-|x-y|^{2}}f(y)dy\right\vert \leq \Vert f\Vert
_{\infty }\int_{B_{R_{0}}}e^{-|x-y|^{2}}dy \\ 
\\ 
\leq \Vert f\Vert _{\infty }\int_{B_{R_{0}}}e^{-(r-R_{0})^{2}}dy=\Vert
f\Vert _{\infty }~|B_{R_{0}}|e^{-(r-R_{0})^{2}}~,%
\end{array}
\label{eqTheoremConvolution6}
\end{equation}%
where $|B_{R_{0}}|$ denotes the Lebesgue measure of the ball of radius $%
R_{0} $. Next, fix $\epsilon >0$ and set $\Delta _{x}=r-(R_{0}+\epsilon )$.
We shall assume henceforth that $r>R_{0}+3\epsilon $, define 
\begin{equation}
x_{0}=\frac{x}{r}(R_{0}+2\epsilon )  \label{eqTheoremConvolution7}
\end{equation}%
and consider the balls $B_{\Delta _{x}}(x)=\left\{ y\in \mathbb{R}%
^{n}:~|x-y|\leq \Delta _{x}\right\} $ and $B_{\epsilon }(x_{0})=\left\{ y\in 
\mathbb{R}^{n}:~|x_{0}-y|\leq \epsilon \right\} $. If $B_{R_{0}}^{c}$ is the
complement of the ball $B_{R_{0}}$, then we have the obvious inclusions: 
\begin{equation}
B_{\epsilon }(x_{0})\subset B_{\Delta _{x}}(x)\subset B_{R_{0}}^{c}~.
\label{eqTheoremConvolution8}
\end{equation}%
Moreover 
\begin{equation}
|x-y|\geq r-R_{0}~,~\forall y\in B_{R_{0}}~.  \label{eqTheoremConvolution9}
\end{equation}%
If we set $M_{\epsilon ,R_{0}}=\text{min}\left\{ f(y):~y\in A_{\epsilon
,R_{0}}(x_{0})\right\} $, where $A_{\epsilon ,R_{0}}(x_{0})$ is the
`annulus' 
\begin{equation}
A_{\epsilon ,R_{0}}(x_{0})=\left\{ y:~R_{0}+\epsilon \leq |y|\leq
R_{0}+3\epsilon \right\} \text{,}  \label{eqTheoremConvolution10}
\end{equation}%
then we have from (\ref{eqTheoremConvolution7})-(\ref{eqTheoremConvolution10}%
): 
\begin{equation}
\begin{array}{c}
\int_{B_{R_{0}}^{c}}f(y)e^{-|x-y|^{2}}dy\geq \int_{B_{\Delta
_{x}}(x)}f(y)e^{-|x-y|^{2}}dy \\ 
\\ 
\geq e^{-\Delta _{x}^{2}}\int_{B_{\Delta _{x}}(x)}f(y)dy\geq e^{-\Delta
_{x}^{2}}\int_{B_{\epsilon }(x_{0})}f(y)dy \\ 
\\ 
\geq e^{-\Delta _{x}^{2}}M_{\epsilon ,R_{0}}|B_{\epsilon
}(x_{0})|=S_{\epsilon ,R_{0}}~e^{-r^{2}+2(R_{0}+\epsilon )r-(R_{0}+\epsilon
)^{2}}~,%
\end{array}
\label{eqTheoremConvolution11}
\end{equation}%
where $S_{\epsilon ,R_{0}}=M_{\epsilon ,R_{0}}|B_{\epsilon }(x_{0})|$ is a
positive constant, which depends only on $R_{0},\epsilon $ and the dimension 
$n$, and where we used the inclusion $B_{\epsilon }(x_{0})\subset
A_{\epsilon ,R_{0}}(x_{0})$.

\textbf{Step 3. }Altogether, from (\ref{eqTheoremConvolution6}) and (\ref%
{eqTheoremConvolution11}), we obtain: 
\begin{equation}
\begin{array}{c}
\Sigma
(x)=\int_{B_{R_{0}}}e^{-|x-y|^{2}}f(y)dy+%
\int_{B_{R_{0}}^{c}}e^{-|x-y|^{2}}f(y)dy \\ 
\\ 
\geq S_{\epsilon ,R_{0}}~e^{-r^{2}+2(R_{0}+\epsilon )r-(R_{0}+\epsilon
)^{2}}-\Vert f\Vert _{\infty }~|B_{R_{0}}|e^{-(r-R_{0})^{2}}= \\ 
\\ 
=e^{-(r-R_{0})^{2}}\left[ \frac{S_{\epsilon ,R_{0}}e^{2\epsilon r}}{%
e^{\epsilon ^{2}+2\epsilon R_{0}}}-\Vert f\Vert _{\infty }~|B_{R_{0}}|\right]
\text{.}%
\end{array}
\label{eqTheoremConvolution12}
\end{equation}%
Again, we have that $\Sigma (x)>0$ for all $\left\vert x\right\vert >R$,
where 
\begin{equation}
R=\text{max}\left\{ R_{0}+3\epsilon ,R_{0}+\frac{\epsilon }{2}+\frac{1}{%
2\epsilon }\ln \left( \frac{\Vert f\Vert _{\infty }~|B_{R_{0}}|}{S_{\epsilon
,R_{0}}}\right) \right\} \text{.}  \label{eqTheoremConvolution13}
\end{equation}

\subsection{Proof of Theorem \protect\ref{TheoremSmootDecay}}

Sufficiency is obvious. Let us prove necessity. Suppose that $Wf$ is such
that its nodal set is bounded. Consider its Husimi function (\ref{eqHusimi1}%
). From Theorem \ref{TheoremConvolution} the nodal set of $Hf(z)$ is also
bounded. In view of (\ref{eqHusimi2}) the Bargmann transform $\mathcal{B}%
f(\zeta )$ is an entire function with a bounded nodal set. By the
Weierstrass factorization theorem it follows that 
\begin{equation}
\mathcal{B}f(\zeta )=Q(\zeta )e^{h(\zeta )}\text{,}
\label{eqProofMainTheorem1}
\end{equation}%
where $Q(\zeta )$ is an entire polynomial of finite degree and $h(\zeta )$
is an entire function. Since $\mathcal{B}f$ has order of growth at most $2$, 
$h$ is a polynomial of degree at most $2$: 
\begin{equation}
h(\zeta )=\alpha \zeta ^{2}+\beta \zeta+\gamma \text{,}
\label{eqProofMainTheorem2}
\end{equation}%
for some $\alpha ,\beta ,\gamma \in \mathbb{C}$. Thus: 
\begin{equation}
\mathcal{B}f(\zeta )=\frac{1}{\sqrt[4]{\pi \hbar }}\int_{\mathbb{R}}e^{-%
\frac{\zeta ^{2}}{2\hbar }-\frac{\sqrt{2}}{\hbar }x\zeta -\frac{x^{2}}{%
2\hbar }}f(x)dx=Q(\zeta )e^{h(\zeta )}\text{.}  \label{eqProofMainTheorem3}
\end{equation}%
Setting $\zeta =ip$ yields: 
\begin{equation}
\begin{array}{c}
\mathcal{B}f(ip)=\frac{1}{\sqrt[4]{\pi \hbar }}\int_{\mathbb{R}}e^{\frac{%
p^{2}}{2\hbar }-\frac{i\sqrt{2}}{\hbar }xp-\frac{x^{2}}{2\hbar }%
}f(x)dx=Q(ip)e^{-\alpha p^{2}+i\beta p+\gamma } \\ 
\\ 
\Leftrightarrow e^{-\frac{p^{2}}{2\hbar }}\mathcal{B}f(ip)=\sqrt{2\pi \hbar }%
\left( \mathcal{F}_{\hbar }(fh_{0})\right) (\sqrt{2}p)=Q(ip)e^{-(\alpha +%
\frac{1}{2\hbar })p^{2}+i\beta p+\gamma }\text{.}%
\end{array}
\label{eqProofMainTheorem4}
\end{equation}%
Under the inverse Fourier transform a generalized Gaussian times a
polynomial is again a generalized Gaussian multiplied by a polynomial: 
\begin{equation}
(fh_{0})(x)=q\left( i\hbar \frac{d}{dx}\right) e^{-\frac{a^{\prime }}{2\hbar 
}x^{2}+\frac{b^{\prime }}{\hbar }x+c^{\prime }}\text{,}
\label{eqProofMainTheorem5}
\end{equation}%
for some polynomial $q(x)$. It follows that: 
\begin{equation}
f(x)=p_{N}^{\prime }(x)e^{-\frac{a}{2\hbar }x^{2}+\frac{b}{\hbar }x+c}\text{,%
}  \label{eqProofMainTheorem6}
\end{equation}%
for some $a,b,c\in \mathbb{C}$, with $Re(a)>0$ and where $p_{N}^{\prime }$
is a polynomial of finite degree $N\in \mathbb{N}$. In view of (\ref%
{eqWigner14}) we can rewrite $f$ as: 
\begin{equation}
f(x)=\pi (z_{1})\mu (S^{-1})g_{N}(x)\text{,}  \label{eqProofMainTheorem7}
\end{equation}%
where 
\begin{equation}
g_{N}(x)=p_{N}(x)e^{-\frac{x^{2}}{2\hbar }}\text{.}
\label{eqProofMainTheorem8}
\end{equation}%
Here $p_{N}(x)$ is a polynomial of degree $N$, $S$ is as in (\ref%
{eqWigner16B}), and $z_{1}=Sz_{0}$, where $z_{0}$ is given by (\ref%
{eqWigner16A}). On the other hand, we can write: 
\begin{equation}
p_{N}(x)=\sum_{n=0}^{N}\frac{b_{n}}{\sqrt[4]{\pi \hbar} \sqrt{2^{n}n!}}%
H_{n}\left(\frac{x}{\sqrt{\hbar}} \right)\text{,}
\label{eqProofMainTheorem10}
\end{equation}%
for some complex constants $b_{0},b_{1},\cdots ,b_{N}$. Substituting (\ref%
{eqProofMainTheorem8},\ref{eqProofMainTheorem10}) in (\ref%
{eqProofMainTheorem7}), we recover (\ref{eqTheoremEllipse1}). The
corresponding Wigner function is: 
\begin{equation}
Wf(z)=W\left( \pi (z_{1})\mu (S^{-1})g_{N}\right) (z)=Wg_{N}(S(z-z_{1}))%
\text{.}  \label{eqProofMainTheorem9}
\end{equation}

\subsection{Proof of Corollary \protect\ref{CorollaryUniquenessSet}}

Suppose that the Wigner functions $Wf$ and $Wg$ have bounded nodal sets.
Then the functions $f$ and $g$ are of the form 
\begin{equation}
f=\sum_{n=0}^{N}b_{n}\pi (z_{1})\mu (S_{1}^{-1})h_{n}~,\hspace{0.5cm}%
g=\sum_{n=0}^{M}c_{n}\pi (z_{2})\mu (S_{2}^{-1})h_{n}~.
\label{eqCorollaryUniquenessSet1}
\end{equation}%
and the associated Wigner functions given by: 
\begin{equation}
Wf(z)=P_{2N}(z)e^{-\frac{1}{\hbar }(z-z_{1})\cdot M_{1}(z-z_{1})}~,\hspace{%
0.5cm}Wg(z)=Q_{2M}(z)e^{-\frac{1}{\hbar }(z-z_{2})\cdot M_{2}(z-z_{2})}~,
\label{eqCorollaryUniquenessSet2}
\end{equation}%
where $M_{1}=S_{1}^{T}S_{1}$, $M_{2}=S_{2}^{T}S_{2}$, and $P_{2N}(z)$ and $%
Q_{2M}(z)$ are polynomials of degree $2N$ and $2M$, respectively. Suppose
that 
\begin{equation}
Wf(z)=Wg(z)~,  \label{eqCorollaryUniquenessSet3}
\end{equation}%
for all $z$ in some open domain $D$. The left-hand side of (\ref%
{eqCorollaryUniquenessSet3}) has terms of the form $x^{n}p^{m}e^{-\frac{1}{%
\hbar }(z-z_{1})\cdot M_{1}(z-z_{1})}$ for $n+m\leq 2N$, whereas the
right-hand side has terms of the form $x^{j}p^{k}e^{-\frac{1}{\hbar }%
(z-z_{2})\cdot M_{2}(z-z_{2})}$ for $j+k\leq 2M$. These terms are all
linearly independent on any domain $D$, except if $z_{1}=z_{2}$, $%
M_{1}=M_{2} $, in which case we also have $N=M$ and $P_{2N}(z)=P_{2M}(z)$ in 
$D$. If the polynomials are equal in $D$, then they are equal in $\mathbb{R}%
^{2}$, which proves the result.

\subsection{Proof of Theorem \protect\ref{TheoremEllipse}}

It follows from (\ref{eqProofMainTheorem9}) that 
\begin{equation}
Wf(z)=Wg_{N}(S_{1}(z-z_{1}^{\prime }))~,  \label{eqProofMainTheorem11AB}
\end{equation}%
where $S_{1}$ is a symplectic matrix, $z_{1}^{\prime }\in \mathbb{R}^{2}$
and $Wg_{N}$ is 
\begin{equation}
\begin{array}{c}
Wg_{N}(z)=\sum_{n,m=0}^{N}b_{n}\overline{b_{m}}W(h_{n},h_{m})(z)= \\ 
\\ 
=\sum_{n=0}^{N}|b_{n}|^{2}\frac{(-1)^{n}}{\pi \hbar }e^{-\frac{|z|^{2}}{%
\hbar }}L_{n}\left( \frac{2|z|^{2}}{\hbar }\right) + \\ 
\\ 
+2Re\sum_{n=0}^{N-1}\sum_{k=1}^{N-n}b_{n}\overline{b_{n+k}}\frac{(-1)^{n}}{%
\pi \hbar }\sqrt{\frac{n!}{(n+k)!}}a^{k}e^{-\frac{|z|^{2}}{\hbar }%
}L_{n}^{(k)}\left( \frac{2|z|^{2}}{\hbar }\right) ~.%
\end{array}
\label{eqProofMainTheorem11}
\end{equation}%
Now, suppose first that $Wf$ vanishes on a circle $|z|=R$. Set $a=\sqrt{%
\frac{2}{\hbar }}Re^{i\theta }$, $0\leq \theta <2\pi $, or equivalently $%
z=R(\cos (\theta ),\sin (\theta ))$. It follows from (\ref%
{eqProofMainTheorem11AB}) that: 
\begin{equation}
\begin{array}{c}
\sum_{n=0}^{N}|b_{n}|^{2}(-1)^{n}L_{n}(|A_{R}|^{2})+ \\ 
\\ 
+\sum_{n=0}^{N-1}\sum_{k=1}^{N-n}(-1)^{n}\sqrt{\frac{n!}{(n+k)!}}%
L_{n}^{(k)}(|A_{R}|^{2})\left( b_{n}\overline{b_{n+k}}A_{R}^{k}+\overline{%
b_{n}}b_{n+k}\overline{A_{R}}^{k}\right) =0~,%
\end{array}
\label{eqProofMainTheorem11A}
\end{equation}%
where 
\begin{equation}
A_{R}=\beta _{R}e^{i\theta }+\gamma _{R}e^{-i\theta }-A_{1}^{\prime }~.
\label{eqProofMainTheorem11B1}
\end{equation}%
The parameters $\beta _{R}$,$\gamma _{R}$ and $A_{1}^{\prime }$ are given
by: 
\begin{equation}
\left\{ 
\begin{array}{l}
\beta _{R}=\frac{R}{\sqrt{2\hbar }}(a_{1}+d_{1}+i(c_{1}-b_{1})) \\ 
\\ 
\gamma _{R}=\frac{R}{\sqrt{2\hbar }}(a_{1}-d_{1}+i(c_{1}+b_{1})) \\ 
\\ 
A_{1}^{\prime }=\sqrt{\frac{2}{\hbar }}\left( a_{1}x_{1}^{\prime
}+b_{1}p_{1}^{\prime }+i(c_{1}x_{1}^{\prime }+d_{1}p_{1}^{\prime })\right) ~,%
\end{array}%
\right.  \label{eqProofMainTheorem11C1}
\end{equation}%
where $z_{1}^{\prime }=(x_{1}^{\prime },p_{1}^{\prime })$, and $%
a_{1},b_{1},c_{1},d_{1}$ are the entries of the symplectic matrix $S_{1}$: 
\begin{equation}
S_{1}=\left( 
\begin{array}{cc}
a_{1} & b_{1} \\ 
c_{1} & d_{1}%
\end{array}%
\right)  \label{eqProofMainTheorem11D1}
\end{equation}%
The trigonometric polynomials $e^{i\theta k}$, $k\in \mathbb{Z}$ are
linearly independent for $\theta \in \left[ 0,2\pi \right] $. The highest
such polynomial in (\ref{eqProofMainTheorem11A}) is $e^{2i\theta N}$.
Setting the associated coefficient to zero yields: 
\begin{equation}
|b_{N}|^{2}(\beta _{R}\overline{\gamma _{R}})^{N}=0~.
\label{eqProofMainTheorem11E1}
\end{equation}%
Since $b_{N}\neq 0$, we must have either $\beta _{R}=0$ or $\gamma _{R}=0$.
Suppose $\beta _{R}=0$. From (\ref{eqProofMainTheorem11C1}): 
\begin{equation}
d_{1}=-a_{1}~,~\text{ and }~c_{1}=b_{1}~.  \label{eqProofMainTheorem11F1}
\end{equation}%
Consequently: 
\begin{equation}
S_{1}=\left( 
\begin{array}{cc}
a_{1} & b_{1} \\ 
b_{1} & -a_{1}%
\end{array}%
\right)  \label{eqProofMainTheorem11G1}
\end{equation}%
But since $S_{1}$ is symplectic, $1=\det (S_{1})=-a_{1}^{2}-b_{1}^{2}$,
which is impossible. Thus we must have $\gamma _{R}=0$: 
\begin{equation}
S_{1}=\left( 
\begin{array}{cc}
a_{1} & b_{1} \\ 
-b_{1} & a_{1}%
\end{array}%
\right) ~,  \label{eqProofMainTheorem11H1}
\end{equation}%
with $a_{1}^{2}+b_{1}^{2}=1$. Hence, we can parametrize $S_{1}$ as: 
\begin{equation}
S_{1}=\left( 
\begin{array}{cc}
\cos (\alpha ) & \sin (\alpha ) \\ 
-\sin (\alpha ) & \cos (\alpha )%
\end{array}%
\right) ~,  \label{eqProofMainTheorem11I2}
\end{equation}%
for some $0\leq \alpha <2\pi $. From (\ref{eqProofMainTheorem11I2}) we
realize that $S_{1}$ is also an orthogonal matrix.

To proceed let us make the following observation:

\begin{lemma}
\label{LemmaOrthosymplectic} Let $S_{1}\in O(2;\mathbb{R})$. Thus, $S_{1}$
admits a parametrization of the form (\ref{eqProofMainTheorem11I2}) for some 
$0\leq \alpha <2\pi $. Suppose that 
\begin{equation}
Wg_{N}(z)=\sum_{n,m=0}^{N}b_{n}\overline{b_{m}}W(h_{n},h_{m})(z)\text{.}
\label{eqLemmaOrthosymplectic1}
\end{equation}%
Then, 
\begin{equation}
Wg_{N}\left( S_{1}(z-z_{1}^{\prime })\right) =\sum_{n,m=0}^{N}c_{n}\overline{%
c_{m}}W(h_{n},h_{m})(z-z_{1}^{\prime })\text{,}
\label{eqLemmaOrthosymplectic1}
\end{equation}%
where the constants are related by $c_{n}=b_{n}e^{i\alpha n}$.
\end{lemma}

\begin{proof}
Let $a = \sqrt{\frac{2}{\hbar}}(x+ip)$. Under the transformation $S_1$ given by (\ref{eqProofMainTheorem11I2}), we have $a \mapsto e^{-i \alpha} a$. If we plug this in (\ref{eqProofMainTheorem11}), the result follows.
\end{proof}

Resuming the proof of Theorem \ref{TheoremEllipse}, the previous Lemma
yields 
\begin{equation}
Wf(z)=\sum_{n,m=0}^{N}c_{n}\overline{c_{m}}W(h_{n},h_{m})(z-z_{1}^{\prime
})~.  \label{eqProofMainTheorem11K1}
\end{equation}%
Next suppose that $Wf$ vanishes on the ellipse 
\begin{equation}
(z-z_{0})\cdot M(z-z_{0})=1~.  \label{eqProofMainTheorem11L1}
\end{equation}%
By Williamson's Theorem \cite{Williamson} there exists a (non-unique)
symplectic matrix $S$, such that (\ref{eqTheoremEllipse2}) holds. Thus, the
Wigner function 
\begin{equation}
Wu(z)=Wf(S^{-1}z+z_{0})  \label{eqProofMainTheorem11N1}
\end{equation}%
vanishes on the circle $|z|=\frac{1}{\sqrt[4]{\det M}}$. Hence $Wu$ must be
of the form (\ref{eqProofMainTheorem11K1}) and (\ref{eqTheoremEllipse1}), (%
\ref{eqTheoremEllipse3}) follow with $z_{1}=z_{0}+S^{-1}z_{1}^{\prime }$.

\subsection{Proof of Corollary \protect\ref{CorollaryEllipse}}

To prove this Corollary, we need the following Lemma.

\begin{lemma}
\label{LemmaLinearIndependence} Let $z_{1},z_{2}\in \mathbb{R}^{2}$, $%
S_{1},S_{2}\in Sp(2;\mathbb{R})$, and $N,M\in \mathbb{N}$. Then the set of
functions 
\begin{equation}
\left\{ \pi (z_{1})\mu (S_{1})h_{0},...,\pi (z_{1})\mu (S_{1})h_{N},\pi
(z_{2})\mu (S_{2})h_{0},...,\pi (z_{2})\mu (S_{2})h_{M}\right\}
\label{eqLemmaLinearIndependence1}
\end{equation}%
are linearly independent, unless 
\begin{equation}
S_{1}S_{1}^{T}=S_{2}S_{2}^{T}~\text{ and }~z_{1}=z_{2}~.
\label{eqLemmaLinearIndependence1A}
\end{equation}
\end{lemma}

\begin{proof}
Suppose that 
\begin{equation}
\sum_{n=0}^Nb_n \pi(z_1)\mu(S_1) h_n+\sum_{m=0}^M c_m \pi(z_2)\mu(S_2)
h_m=0~,  \label{eqLemmaLinearIndependence2}
\end{equation}
for some constants $(b_0,\cdots,b_N,c_0, \cdots, c_M) \in \mathbb{C}^{N+M+2}$.

From (\ref{eqWigner4A}), (\ref{eqWigner4B}) and (\ref{eqWigner9}), we have: 
\begin{equation}
\sum_{n=0}^{N}b_{n}h_{n}=\sum_{m=0}^{M}d_{m}\pi (z_{0})\mu (S)h_{m}~,
\label{eqLemmaLinearIndependence3}
\end{equation}%
where 
\begin{equation}
d_{m}=-c_{m}e^{-\frac{i}{\hbar }%
x_{1}(p_{1}-p_{2})}~,~z_{0}=S_{1}^{-1}(z_{2}-z_{1})~,~S=S_{1}^{-1}S_{2}~.
\label{eqLemmaLinearIndependence4}
\end{equation}%
Equation (\ref{eqLemmaLinearIndependence3}) can be written in terms of the
Wigner distributions as: 
\begin{equation}
\sum_{n,j=0}^{N}b_{n}\overline{b_{j}}W(h_{n},h_{j})(z)=\sum_{m,k=0}^{M}d_{m}%
\overline{d_{k}}W(h_{m},h_{k})(S^{-1}(z-z_{0}))~,
\label{eqLemmaLinearIndependence5}
\end{equation}%
for all $z\in \mathbb{R}^{2}$. If we write $z_{0}=(x_{0},p_{0})$ and 
\begin{equation}
(SS^{T})^{-1}=\left( 
\begin{array}{cc}
a & b \\ 
b & c%
\end{array}%
\right) ~,  \label{eqLemmaLinearIndependence6}
\end{equation}%
with $a,c>0$, and $ac-b^2=1$, then (\ref{eqLemmaLinearIndependence5}) can be
re-expressed as: 
\begin{equation}
e^{-\frac{1}{\hbar }(x^{2}+p^{2})}P_{2N}(x,p)=e^{-\frac{1}{\hbar }\left(
a(x-x_{0})^{2}+2b(x-x_{0})(p-p_{0})+c(p-p_{0})^{2}\right) }Q_{2M}(x,p)~,
\label{eqLemmaLinearIndependence7}
\end{equation}%
where $P_{2N}$ and $Q_{2M}$ are polynomials of degree $2N$ and $2M$,
respectively. Suppose that the constants $b_{0},\cdots ,b_{N},d_{0},\cdots
,d_{M}$ in (\ref{eqLemmaLinearIndependence5})\ do not all vanish. A
straightforward asymptotic analysis shows that this can only happen
provided: 
\begin{equation}
SS^{T}=I~,~z_{0}=0.  \label{eqLemmaLinearIndependence8}
\end{equation}%
This means that $S\in Sp(2;\mathbb{R})\cap O(2)$, and from (\ref%
{eqLemmaLinearIndependence4}): 
\begin{equation}
S_{1}S_{1}^{T}=S_{2}S_{2}^{T}~,z_{1}=z_{2}~.
\label{eqLemmaLinearIndependence9}
\end{equation}
\end{proof}

We are now in a position of concluding the proof of Corollary \ref%
{CorollaryEllipse}. Suppose that $\mathcal{N}(Wf)$ is bounded and contains
two ellipses: $(z-z_{1})\cdot M_{1}(z-z_{1})=1$ and $(z-z_{2})\cdot
M_{2}(z-z_{2})=1$. From the previous results $f$ must be of the form: 
\begin{equation}
f=\sum_{n=0}^{N}b_{n}\pi (z_{1}^{\prime })\mu (S_{1}^{-1})h_{n}~,
\label{eqConcentriEllipses1}
\end{equation}%
but also: 
\begin{equation}
f=\sum_{m=0}^{M}c_{m}\pi (z_{2}^{\prime })\mu (S_{2}^{-1})h_{m}~,
\label{eqConcentriEllipses2}
\end{equation}%
for some $z_{1}^{\prime },z_{2}^{\prime }\in \mathbb{R}^{2}$, $(b_{0},\cdots
,b_{N},c_{0},\cdots ,c_{M})\in \mathbb{C}^{N+M+2}$ and symplectic matrices $%
S_{1}$ and $S_{2}$, such that 
\begin{equation}
M_{1}=\sqrt{\det M_{1}}~S_{1}^{T}S_{1}~\text{ and }M_{2}=\sqrt{\det M_{2}}%
~S_{2}^{T}S_{2}~  \label{eqConcentriEllipses3}
\end{equation}%
Equating (\ref{eqConcentriEllipses1}) and (\ref{eqConcentriEllipses2})
yields: 
\begin{equation}
\sum_{n=0}^{N}b_{n}\pi (z_{1}^{\prime })\mu
(S_{1}^{-1})h_{n}-\sum_{m=0}^{M}c_{m}\pi (z_{2}^{\prime })\mu
(S_{2}^{-1})h_{m}=0~.  \label{eqConcentriEllipses4}
\end{equation}%
If $f$ is not identically zero, then by Lemma \ref{LemmaLinearIndependence}
this is possible only if $z_{1}^{\prime}=z_{2}^{\prime}$ and $%
S_{1}^{T}S_{1}=S_{2}^{T}S_{2}$. By (\ref{eqConcentriEllipses3}), we conclude
that there exists $\mu >0$ such that $M_{2}=\mu M_{1}$ and the result
follows.

\subsection{Proof of Theorem \protect\ref{TheoremLaguerreDivisibility}}

Before we embark on the proof let us quickly revise some necessary notation.

We use the familiar notation in modular arithmetic $a=b$ (mod $n$) to mean
that two integers $a$ and $b$ are congruent \textit{modulo} the integer $n$.
We write $a|b$ if $a$ divides $b$. Given a ring $\mathbb{F}$ we denote by $%
\mathbb{F} \left[x \right]$ the set of polynomials in the variable $x$ with
coefficients in $\mathbb{F}$. A polynomial $P(x) \in \mathbb{F} \left[x %
\right]$ is said to be monic if the coefficient of the highest power is
exactly one. Given two polynomials $P(x)$ and $Q(x)$ we denote by gcd$(P,Q)$
their greatest common divisor, that is the polynomial of highest degree that
divides both $P$ and $Q$.

For a prime $p$ we denote by $\mathbb{F}_p= \mathbb{Z}/ \mathbb{Z}%
_p=\left\{0,1, \cdots, p-1 \right\}$ the finite field of order $p$ with
addition and multiplication performed \textit{modulo} $p$. The reduction map 
$\pi_p: \mathbb{Z} \left[x \right] \to \mathbb{F}_p \left[x \right]$, $P(x)
\mapsto \overline{P}(x)$ reduces each coefficient of $P$ \textit{modulo} $p$%
. If $P(x)|Q(x)$ in $\mathbb{Z} \left[x \right] $, then $\overline{P}(x)|%
\overline{Q}(x)$ in $\mathbb{F}_p \left[x \right] $ for any prime $p$
(Chapter IV \cite{Lang}).

\vspace{0.3 cm} First, let us make the following brief remark.

\begin{remark}
\label{RemarkLaguerreDivisibility} Notice that, in terms of Laguerre
polynomials, Theorem \ref{TheoremLaguerreDivisibility} is all that we need
for our proofs. Nevertheless, one could be tempted to extend this and
consider the divisibility of $L_n^{(m)}$ by $L_k^{(j)}$ with $j \geq 1$.
Indeed, one can show, with some minor modifications of our proof, that (i)
if $n < 2k$, this is possible if and only $L_n^{(m)}=L_k^{(j)}$, and (ii)
that this is impossible if $n \geq 2k +2j$. However, in the range $2k \leq n
< 2k+2j$, one can find counterexamples. For instance: 
\begin{equation*}
\begin{array}{l}
L_2^{(2)}(x)=- \frac{1}{2}(x-6) L_1^{(1)} (x) \\ 
\\ 
L_4^{(5)}(x)=- \frac{1}{24} \left(x^3-30x^2+252x-504 \right) L_1^{(5)}(x)~.%
\end{array}%
\end{equation*}
\end{remark}

\vspace{0.3 cm} Now we address the proof of Theorem \ref%
{TheoremLaguerreDivisibility}.

Notice that $\text{deg}(L_k)=k$ and $\text{deg}(L_n^{(m)})=n$. Thus, if (\ref%
{eqTheoremLaguerreDivisibility1}) holds, then we must have: 
\begin{equation}
n \geq k~.  \label{eqProofTheoremLaguerreDivisibility1}
\end{equation}

The proof will proceed in two steps.

\begin{enumerate}
\item[\textbf{Step 1:}] We assume $n < 2k$. Set $P(x)=L_k(x)$ and $%
Q(x)=L_n^{(m)}(x)$. Since $P|Q$ we can write 
\begin{equation}
Q(x)=P(x)R(x)~,  \label{eqProofTheoremLaguerreDivisibility2}
\end{equation}
where $R(x)$ is a polynomial of degree $l:=\text{deg}(R)=n-k$. From (\ref%
{eqLaguerreDiffEq}) we have: 
\begin{equation}
\begin{array}{l}
xP^{\prime\prime} +(1-x) P^{\prime}+k P=0~, \\ 
\\ 
xQ^{\prime\prime} +(m+1-x) Q^{\prime}+n Q=0~.%
\end{array}
\label{eqProofTheoremLaguerreDivisibility3}
\end{equation}
Plugging $Q=PR$ is the last equation we obtain: 
\begin{equation}
x(P^{\prime\prime} R +2 P^{\prime}R^{\prime}+ PR^{\prime\prime})+
(m+1-x)(P^{\prime}R+PR^{\prime})+ nPR=0~.
\label{eqProofTheoremLaguerreDivisibility4}
\end{equation}
From the first equation in (\ref{eqProofTheoremLaguerreDivisibility3}) it
then follows that: 
\begin{equation}
xPR^{\prime\prime} +\left(2xP^{\prime} + (m+1-x)P \right) R^{\prime} +(m
P^{\prime} +l P)R=0~.  \label{eqProofTheoremLaguerreDivisibility5}
\end{equation}
Reducing this equation \textit{modulo} $P$ yields: 
\begin{equation}
P^{\prime} (2x R^{\prime}+mR) \equiv 0 \hspace{0.3 cm} (\text{mod }P)
\label{eqProofTheoremLaguerreDivisibility6}
\end{equation}
Since the polynomial $P=L_k$ has only simple roots, we have gcd$%
(P,P^{\prime})=1$. Therefore 
\begin{equation}
P|(2x R^{\prime}+mR)~.  \label{eqProofTheoremLaguerreDivisibility7}
\end{equation}
By assumption $n < 2k$. Thus: 
\begin{equation}
\text{deg}(2x R^{\prime}+mR) \leq \text{deg}(R)=l=n-k < k=\text{deg}(P)~.
\label{eqProofTheoremLaguerreDivisibility8}
\end{equation}
This forces 
\begin{equation}
2x R^{\prime}+mR \equiv 0~.  \label{eqProofTheoremLaguerreDivisibility9}
\end{equation}
Consequently, $R$ is of the form $R(x)=C x^{-m/2}$ for some constant $C$.
This is a polynomial only if $m=0$ and $R$ is constant. Therefore $l=n-k=0$.
Altogether $L_k=L_n^{(m)}$.

\item[\textbf{Step 2:}] We now consider the case $n\geq 2k$. For $n,m\in 
\mathbb{N}_{0}$ we observe that 
\begin{equation}
A_{n}^{(m)}(x):=(-1)^{n}n!L_{n}^{(m)}(x)=\sum_{j=0}^{n}(-1)^{j}\frac{n!}{%
(n-j)!}\left( 
\begin{array}{c}
n+m \\ 
j%
\end{array}%
\right) x^{n-j}~  \label{eqProofTheoremLaguerreDivisibility10}
\end{equation}
is a monic polynomial of degree $n$ with integer coefficients. Let us also
define 
\begin{equation}
A_{k}(x):=A_{k}^{(0)}(x)~.  \label{eqProofTheoremLaguerreDivisibility11}
\end{equation}%
Notice that $A_{k}$ satisfies: 
\begin{equation}
A_{k}(0)=(-1)^{k}k!~.  \label{eqProofTheoremLaguerreDivisibility12}
\end{equation}

\begin{lemma}
\label{LemmaProofTheoremLaguerreDivisibility1} If $L_k| L_n^{(m)}$ in $%
\mathbb{C}\left[x \right]$, then $A_k| A_n^{(m)}$ in $\mathbb{Z}\left[x %
\right]$.
\end{lemma}

\begin{proof}
If $L_k| L_n^{(m)}$ in $\mathbb{C}\left[x \right]$, then $A_k| A_n^{(m)}$ in $\mathbb{C}\left[x \right]$. Since both polynomials have integer coefficients and $A_k$ is monic, polynomial division over $\mathbb{Z}$ gives zero remainder and a quotient in $\mathbb{Z}\left[x \right]$, according to the division algorithm for monic polynomials over a commutative ring (see Theorem 1.1, Ch.IV in \cite{Lang}).
\end{proof}

\begin{lemma}
\label{LemmaProofTheoremLaguerreDivisibility2} Let $0 \leq r <n$, and let $p$
be a prime number such that 
\begin{equation}
p|(n-r)~, \hspace{0.5 cm} p>r~.  \label{eqProofTheoremLaguerreDivisibility13}
\end{equation}
Then, 
\begin{equation}
A_n^{(m)}(x)=x^{n-r} A_r^{(m)}(x) \hspace{0.3 cm} (\text{mod }p)~.
\label{eqProofTheoremLaguerreDivisibility14}
\end{equation}
\end{lemma}

\begin{proof}
Consider the coefficient of $x^{n-j}$ in (\ref{eqProofTheoremLaguerreDivisibility10}). If $j >r$, then
\begin{equation}
\frac{n!}{(n-j)!}=n(n-1) \cdots (n-j+1)
\label{eqProofTheoremLaguerreDivisibility15}
\end{equation}
contains the factor $(n-r)$, so it is divisible by $p$. Hence the coefficient of $x^{n-j}$ in $A_n^{(m)}$ is zero \textit{modulo} $p$.

On the other hand, if $0 \leq j \leq r$, then $j <p$ and $n \equiv r$ (mod $p$). Hence
\begin{equation}
\frac{n!}{(n-j)!} \equiv \frac{r!}{(r-j)!} \hspace{0.3 cm} (\text{mod }p)~.
\label{eqProofTheoremLaguerreDivisibility16}
\end{equation}
And, since $j!$ is invertible \textit{modulo} p,
\begin{equation}
\left(
\begin{array}{c}
n+m\\
j
\end{array}
\right) \equiv \left(
\begin{array}{c}
r+m\\
j
\end{array}
\right) \hspace{0.3 cm} (\text{mod }p)~.
\label{eqProofTheoremLaguerreDivisibility17}
\end{equation}
Therefore the coefficient of $x^{n-j}$ in $A_n^{(m)}$ is congruent to the coefficient of $x^{n-j}$ in $x^{n-r} A_r^{(m)}$. This proves (\ref{eqProofTheoremLaguerreDivisibility14}). 
\end{proof}
In the sequel we will resort to the Sylvester-Schur theorem (see \cite%
{Erdos,Schur1,Sylvester}).

\begin{lettertheorem}[Sylvester, Schur]
\label{TheoremSylvesterSchur} Let $a,k \in \mathbb{N}$ with $a>k$. Then the
product 
\begin{equation*}
a(a+1) \cdots (a+k-1)
\end{equation*}
has a prime divisor greater than $k$.
\end{lettertheorem}

We are now in a position to prove that $L_k |L_n^{(m)}$ is impossible if $%
n\geq 2k$.

Indeed, suppose that $L_k |L_n^{(m)}$ and $n\geq 2k$. By Lemma \ref%
{LemmaProofTheoremLaguerreDivisibility1} 
\begin{equation}
A_k |A_n^{(m)}  \label{eqProofTheoremLaguerreDivisibility18}
\end{equation}
in $\mathbb{Z} \left[x \right]$. Let us apply Theorem \ref%
{TheoremSylvesterSchur} to $a=n-k+1$. Since $n \geq 2k$, we have $a \geq
k+1>k$. Therefore the product 
\begin{equation*}
(n-k+1)(n-k+2) \cdots n
\end{equation*}
has a prime divisor $p>k$. Hence for some $r \in \left\{0,1, \cdots, k-1
\right\}$, 
\begin{equation}
p|(n-r)~.  \label{eqProofTheoremLaguerreDivisibility19}
\end{equation}
Since $p>k>r$, Lemma \ref{LemmaProofTheoremLaguerreDivisibility2} gives 
\begin{equation}
A_n^{(m)}(x) \equiv x^{n-r} A_r^{(m)}(x) \hspace{0.5 cm} (\text{mod } p)
\label{eqProofTheoremLaguerreDivisibility20}
\end{equation}
Reducing (\ref{eqProofTheoremLaguerreDivisibility18}) \textit{modulo} $p$
and using (\ref{eqProofTheoremLaguerreDivisibility20}) we get 
\begin{equation}
\overline{A_k}| x^{n-r} \overline{A_r^{(m)}}
\label{eqProofTheoremLaguerreDivisibility21}
\end{equation}
in $\mathbb{F}_p \left[x \right]$.

But (\ref{eqProofTheoremLaguerreDivisibility12}) and $p>k$ imply 
\begin{equation*}
\overline{A_k}(0)=(-1)^k k! \not\equiv 0 \hspace{0.5 cm} (\text{mod } p)~.
\end{equation*}
Therefore gcd$(\overline{A_k},x)=1$ in $\mathbb{F}_p \left[x \right]$, and
we may cancel the power of $x$ from the divisibility: 
\begin{equation}
\overline{A_k}| \overline{A_r^{(m)}}~.
\label{eqProofTheoremLaguerreDivisibility21}
\end{equation}
This is impossible because $\overline{A_k}$ has degree $k$ and $\overline{%
A_r^{(m)}}$ has degree $r<k$. This concludes the proof.
\end{enumerate}

\subsection{Proof of Theorem \protect\ref{MainTheorem}}

Sufficiency is obvious. Let us prove necessity. Parts of the argument will
be used in the proof of other results. We divide it in three steps.

\textbf{Step 1.} If $\mathcal{N}(Wf)=\mathcal{N}(Wh_k)$, then Suppose that $%
Wf$ has a bounded nodal set and vanishes on some circle.

From Theorem \ref{TheoremEllipse}, $f$ must be of the form: 
\begin{equation}
f=\sum_{n=0}^{N}c_{n}\pi (z_{0})h_{n}~.  \label{eqProofMainTheorem11A1}
\end{equation}

Thus $f \in \mathcal{H}_{z_0}^N$ for some $N \in \mathbb{N}$ and some $z_0
\in \mathbb{R}^2$.

From Corollary \ref{CorollarypseudoDisp1}, we conclude that 
\begin{equation}
Wf(z)=K Wg(z) e^{- \frac{1}{\hbar}(z_0^2 -2 z_0 z)}~,  \label{eqpseudoDisp31}
\end{equation}
where $g \in \mathcal{H}_{0}^N$ and $Wg$ also vanishes on the same circle.

It follows that 
\begin{equation}
\begin{array}{c}
Wg(z)=\sum_{n,m=0}^{N}c_{n}\overline{c_{m}}W(h_{n},h_{m})(z)= \\ 
\\ 
=\sum_{n=0}^{N}|c_{n}|^{2}\frac{(-1)^{n}}{\pi \hbar }e^{-\frac{|z|^{2}}{%
\hbar }}L_{n}\left( \frac{2|z|^{2}}{\hbar }\right) + \\ 
\\ 
+2Re\sum_{n=0}^{N-1}\sum_{m=1}^{N-n}c_{n}\overline{c_{n+m}}\frac{(-1)^{n}}{%
\pi \hbar }\sqrt{\frac{n!}{(n+m)!}}a^{m}e^{-\frac{|z|^{2}}{\hbar }%
}L_{n}^{(m)}\left( \frac{2|z|^{2}}{\hbar }\right) ~,%
\end{array}
\label{eqProofMainTheorem11F}
\end{equation}%
where, as before, $a= \sqrt{\frac{2}{\hbar}}(x+ip)$.

Writing (\ref{eqProofMainTheorem11F})\ in polar coordinates and imposing
that $Wg(z)$ vanish on the circle $\frac{2|z|^{2}}{\hbar}=R$, leads to 
\begin{equation}
\begin{array}{c}
\sum_{n=0}^{N}|c_{n}|^{2}(-1)^{n}L_{n}\left( R\right) + \\ 
\\ 
+2Re\sum_{m=1}^{N}\sum_{n=0}^{N-m}c_{n}\overline{c_{n+m}}(-1)^{n}\sqrt{\frac{%
n!}{(n+m)!}} R^{m/2}e^{i\theta m}L_{n}^{(m)}\left( R\right) ~=0~,%
\end{array}
\label{eqProofMainTheorem11H}
\end{equation}%
for $\theta \in \left[\right.0, 2 \pi \left. \right)$.

Since the functions $\{e^{i\theta m}\}_{m=0}^{N}$ are linearly independent,
the coefficients of the trigonometric polynomial (\ref{eqProofMainTheorem11H}%
) must be zero. Thus,

\begin{equation}
\left\{ 
\begin{array}{l}
\sum_{n=0}^{N}|c_{n}|^{2}(-1)^{n}L_{n}\left( R\right) =0~, \\ 
\\ 
\sum_{n=0}^{N-m}c_{n}\overline{c_{n+m}}(-1)^{n}\sqrt{\frac{n!}{(n+m)!}}%
L_{n}^{(m)}\left( R\right) =0~,~m=1,\cdots ,N~.%
\end{array}%
\right.  \label{eqProofMainTheorem15}
\end{equation}%
Now suppose that $\mathcal{N}(Wh_{k})\subset \mathcal{N}(Wf)$ for some $k\in 
\mathbb{N}$. Then, from (\ref{eqProofMainTheorem15}) we have, denoting by $%
\rho _{j}^{(k)}$, $j=1,\cdots ,k$ the $k$ distinct roots of $L_{k}$, that 
\begin{equation}
\left\{ 
\begin{array}{l}
\sum_{n=0}^{N}|c_{n}|^{2}(-1)^{n}L_{n}\left( \rho _{j}^{(k)}\right) =0~, \\ 
\\ 
\sum_{n=0}^{N-m}c_{n}\overline{c_{n+m}}(-1)^{n}\sqrt{\frac{n!}{(n+m)!}}%
L_{n}^{(m)}\left( \rho _{j}^{(k)}\right) =0~,~m=1,\cdots ,N~,%
\end{array}%
\right.  \label{eqProofMainTheorem15a}
\end{equation}%
for all $j=1,\cdots ,k$.

\textbf{Step 2. } We will now prove, by induction, that $c_{n}=0$, for all $%
n=0,1,\cdots ,N-1$. The second equation for $m=N$ implies $c_{0}\overline{%
c_{N}}=0$. Since $c_{N}\neq 0$, we conclude that $c_{0}=0$.

Next, suppose that $c_{i}=0$ for all $i$ up to some $l<N-1$. Then the second
equation in (\ref{eqProofMainTheorem15a}) for $m=N-l-1$ becomes: 
\begin{equation}
\begin{array}{c}
\sum_{n=0}^{l+1}c_{n}\overline{c_{n+N-l-1}}(-1)^{n}\sqrt{\frac{n!}{(n+N-l-1)!%
}}L_{n}^{(N-l-1)}\left( \rho _{j}^{(k)}\right) =0 \\ 
\\ 
\Leftrightarrow c_{l+1}\overline{c_{N}}(-1)^{l+1}\sqrt{\frac{(l+1)!}{N!}}%
L_{l+1}^{(N-l-1)}\left( \rho _{j}^{(k)}\right) =0~,%
\end{array}
\label{eqProofMainTheorem15A}
\end{equation}%
for all $j=1,\cdots ,k$. From Theorem \ref{TheoremLaguerreDivisibility},
there exists $j=1,\cdots ,k$, such that $L_{l+1}^{(N-l-1)}\left( \rho
_{j}^{(k)}\right) \neq 0$, whence $c_{l+1}=0$.

\textbf{Step 3.}

Now, consider the first equation in (\ref{eqProofMainTheorem15a}). Setting $%
c_{0}=\cdots =c_{N-1}=0$, we obtain: 
\begin{equation}
|c_{N}|^{2}(-1)^{N}L_{N}\left( \rho _{j}^{(k)}\right) =0\text{,}
\label{eqProofMainTheorem16}
\end{equation}%
for all $j=1,\cdots ,k$. Since $c_{N}\neq 0$, this is possible iff $%
L_{N}\left( \rho _{j}^{(k)}\right) =0$, for all $j=1,\cdots ,k$, in which
case $L_{N}=L_{k}$, by Theorem \ref{TheoremLaguerreDivisibility}.

\subsection{Proof of Theorem \protect\ref{TheoremRadialZeros}}

If we go back to equation (\ref{eqProofMainTheorem15}) and assume that $%
L_n^{(m)}(R) \neq 0 $, for all $n \in \mathbb{N}$, $m \in \mathbb{N}_0$,
then following the same steps we conclude that $Wf \equiv 0$, which is
impossible. Therefore the radius $R$ of a circular zero must be the root of
some Laguerre polynomial.

\subsection{Proof of Proposition \protect\ref{PropositionLaguerrezeros3}}

Let us write $\bar x$ as an irreducible fraction $\bar x=\frac{a}{b}$ for
some $a,b \in \mathbb{N} $. Next we define the sequence: 
\begin{equation}
v_n^{(m)}:= n! b^n L_n^{(m)} (\bar x)= \sum_{j=0}^n (-1)^j \frac{n!}{j!}
\left( 
\begin{array}{c}
n+m \\ 
n-j%
\end{array}
\right) a^j b^{n-j}~.  \label{eqLaguerre16}
\end{equation}
This is a sequence of integers and we have: 
\begin{equation}
v_n^{(m)}= (-1)^n a^n ~ \hspace{0.5 cm} (\text{mod } b) .
\label{eqLaguerre17}
\end{equation}
Hence, there exists $s \in \mathbb{Z}$ such that 
\begin{equation}
v_n^{(m)}= b s\pm a^n ~.  \label{eqLaguerre18}
\end{equation}
But, since $\frac{a}{b}$ is irreducible, there is no $s \in \mathbb{Z}$ for
which $v_n^{(m)}=0$, unless $b=1$.

Regarding the last statement, we observe that for $\bar x=a$: 
\begin{equation}
v_n^{(m)} = (-1)^n \bar x^n \hspace{0.5 cm} (\text{mod } n)~ ,
\label{eqLaguerre19}
\end{equation}
which means that 
\begin{equation}
v_n^{(m)}= tn \pm \bar x^n ~,  \label{eqLaguerre20}
\end{equation}
for some $t \in \mathbb{Z}$. Hence, $v_n^{(m)}=0$ is possible only if $n$ is
a divisor of $\bar x^n = p_1^{n n_1} \cdots p_k^{n n_k}$, and the result
follows.

\subsection{Proof of Theorem \protect\ref{TheoremRationalRadius}}

Again this follows from the proof of Theorem \ref{MainTheorem} and
Proposition \ref{PropositionLaguerrezeros3}.

For the remaining result, let us assume, without loss of generality, that $%
z_0=0$. Since $\mathcal{N}(Wf)$ is bounded and $Wf$ vanishes on a circle,
from Theorem \ref{TheoremEllipse}, $f$ must be of the form: 
\begin{equation}
f= \sum_{n=0}^N c_n \pi(z_1) h_n~,  \label{eqProofMainTheorem11A1}
\end{equation}
for some $z_1 \in \mathbb{R}^2$. Suppose in addition that (\ref{eqrevision8A}%
) holds.

By Proposition \ref{PropositionBoundWigner}: 
\begin{equation}
f(-x)=\pm f(x)~.  \label{eqProofMainTheorem11B}
\end{equation}
Therefore: 
\begin{equation}
\sum_{n=0}^N c_n \mathcal{I} \pi(z_1) h_n= \pm \sum_{n=0}^N c_n \pi(z_1)
h_n~.  \label{eqProofMainTheorem11C}
\end{equation}
From (\ref{eqWigner13A}) we obtain: 
\begin{equation}
\sum_{n=0}^N c_n \pi(-z_1) \mathcal{I}h_n \mp \sum_{n=0}^N c_n \pi(z_1)
h_n=0~.  \label{eqProofMainTheorem11D}
\end{equation}
Since Hermite functions satisfy $\mathcal{I}h_n=(-1)^n h_n$, it follows
that: 
\begin{equation}
\sum_{n=0}^N c_n (-1)^n \pi(-z_1) h_n \mp \sum_{n=0}^N c_n \pi(z_1) h_n=0~.
\label{eqProofMainTheorem11E}
\end{equation}
By Lemma \ref{LemmaLinearIndependence} this is possible only if $z_1=0$. It
follows that 
\begin{equation}
\sum_{n=0}^N c_n \left[(-1)^n \mp 1\right] h_n=0~.
\label{eqProofMainTheorem11E!}
\end{equation}
Thus, 
\begin{equation}
f= \sum_{n=0}^N c_n h_n~,  \label{eqProofMainTheorem11AF}
\end{equation}
where all coefficients $c_n$ with $n$ odd (resp. even) vanish if $Wf(0)=%
\frac{1}{\pi \hbar}$ (resp. $Wf(0)=-\frac{1}{\pi \hbar}$ )

\begin{enumerate}
\item If $p=2^{k}$, then from Proposition \ref{PropositionLaguerrezeros3} $%
L_{n}^{(m)}(x)$ can vanish if and only if $n=2^{j}$, for some $j \in \mathbb{%
N}$. Thus only the terms with $n$ even survive in (\ref%
{eqProofMainTheorem11AF}). As a result, $Wf(0)=+\frac{1}{\pi \hbar }$.

\item If $p$ is odd, then from Proposition \ref{PropositionLaguerrezeros3} $%
L_{n}^{(m)}(x)$ vanishes only if $n$ is odd. Thus all $h_{n}$ contributing
to the expansion of $f$ are odd, and $Wf(0)=-\frac{1}{\pi \hbar }$.
\end{enumerate}

\subsection{Proof of Theorem \protect\ref{TheoremHermite1}}

\bigskip

Suppose that $Wf$ has a bounded nodal set, and a circular zero of radius $R$%
. After a time-frequency shift we may assume that it vanishes on the circle $%
\frac{2|z|^{2}}{\hbar}=R$. Then $Wf$ is of the form (\ref{eqpseudoDisp31})
where $g \in \mathcal{H}_{0}^N$ and $Wg$ also vanishes on the same circle.

\begin{enumerate}
\item[(1)] Thus, equations (\ref{eqProofMainTheorem15}) are valid. As we
have seen before, these equations have no solution, unless $R$ is a zero of
some of the Laguerre polynomial $L_{n}^{(m)}(x)$ for some pair $(n,m)$ with $%
n =1, \cdots, N$ and $m=0$, or $n=1,\cdots ,N-m$ and $m=1,\cdots ,N$. It is
well known \cite{Gatteschi} that the zeros of the Laguerre polynomial $%
L_{n}^{(m)}(x)$ lie in the oscillatory region $x\in \left( 0,\nu \right) $,
with 
\begin{equation}
\nu =4n+2m+2~.  \label{eqTheoremRank2}
\end{equation}%
We thus have: 
\begin{equation}
R<\nu \leq \text{max} \left\{4N+2, 4N \right\} = 4N+2 ~,
\label{eqTheoremRank3}
\end{equation}%
and we conclude that $N\geq \frac{R-2}{4}$.

\item[(2)] Let us now suppose that $Wf$ vanishes on the circle $\frac{%
2|z|^{2}}{\hbar}=R\leq \frac{1}{k}$.

We prove (\ref{RankN_k}) by contradiction. Let us then assume that $k>N $.
We proceed by induction following the steps of the proof of Theorem \ref%
{MainTheorem}. The second equation in (\ref{eqProofMainTheorem15}) for $m=N$
implies $c_{0}\overline{c_{N}}=0$. Since $c_{N}=\left\langle
f,h_{N}\right\rangle \neq 0$, we conclude that $c_{0}=0$.

Next, assume that $c_{i}=0$ for all $i$ up to some $l<N$. For $\alpha >-1$
and $n\geq 2$, we have the following inequality \cite[(2.9)]{DupTMuld},
where $\rho _{1,\alpha }^{(n)}$ denotes the smallest zero of $L_{n}^{(\alpha
)}(x)$: 
\begin{equation}
\frac{1}{n}<\frac{\rho _{1,\alpha }^{(n)}}{\alpha +1}<\frac{\alpha +2}{%
\alpha +1+n}\text{.}  \label{alpha}
\end{equation}%
Applying this to $\rho _{1}^{(k)}=\rho _{1,0}^{(k)}$ and $\rho
_{1,N-l-1}^{(l+1)}$ with $k\geq N+1$, and $l<N-1$ yields 
\begin{equation}
R\leq \frac{1}{k}<\rho _{1}^{(k)}<\frac{2}{1+k}<\frac{N-l}{l+1}<\rho
_{1,N-l-1}^{(l+1)}\text{,}  \label{0kzeros}
\end{equation}%
From the induction hypothesis and (\ref{eqProofMainTheorem15}), we conclude
that 
\begin{equation}
c_{l+1}\overline{c_{N}}L_{l+1}^{(N-l-1)}(R)=0\text{.}  \label{eqgamma}
\end{equation}%
In view of (\ref{0kzeros}), we have 
\begin{equation}
L_{l+1}^{(N-l-1)}(R)\neq 0,  \label{eqdelta}
\end{equation}%
and since $c_{N}\neq 0$, we must have $c_{l+1}=0$. Consequently, $c_{i}=0$,
for all $i=0,1,\cdots ,N-1$. From the first equation in (\ref%
{eqProofMainTheorem15}), we have $|c_{N}|^{2}L_{N}(R)=0$. But, since $k>N$, (%
\ref{alpha}) and (\ref{0kzeros}) give:%
\begin{equation*}
R\leq \frac{1}{k}<\frac{1}{N}<\rho _{1}^{(N)}\text{.}
\end{equation*}%
Thus, $L_{N}(\rho _{1}^{(k)})\neq 0$, and by (\ref{eqgamma}), $c_{N}=0$. We
conclude that $Wf=0$ identically, which is impossible.
\end{enumerate}

\subsection{Proof of Theorem \protect\ref{TheoremLineSegment}}

If $\mathcal{N}(Wf)$ is bounded, then $Wf$ is of the form 
\begin{equation}
Wf(z)=\sum_{n,m=0}^{N}b_{n}\overline{b_{m}}W(h_{n},h_{m})(S(z-z_{0}))~,
\label{eqProofLineSegment1}
\end{equation}%
where as usual $(b_{0},\cdots ,b_{N})\in \mathbb{C}^{N+1}$, $b_{N}\neq 0$, $%
S\in Sp(2;\mathbb{R})$ and $z_{0}\in \mathbb{R}^{2}$. Since symplectic
transformations map line segments to line segments, we conclude that $%
Wf^{\prime}$, with $f^{\prime}= \mu(S) f \in \mathcal{H}_{z_0}^N$, also
vanishes on a line segment. We can then implement some time frequency shift $%
f^{\prime \prime}= \pi(z_1) f^{\prime} \in \mathcal{H}_{z_0+z_1}^N$ such
that its Wigner function $Wf^{\prime \prime}$ vanishes on a line segment
which has an endpoint located at the origin. Consequently the function $%
g^{\prime}=E_{-z_0-z_1}f^{\prime \prime}$ is such that $g^{\prime} \in 
\mathcal{H}_{0}^N$ and $Wg^{\prime} $ vanishes on the same line segment. It
follows that 
\begin{equation}
Wg^{\prime }(z)=\sum_{n,m=0}^{N}d_{n}\overline{d_{m}}W(h_{n},h_{m})(z)~,
\label{eqProofLineSegment2}
\end{equation}
for some constants $(d_{0},\cdots ,d_{N})\in \mathbb{C}^{N+1}$. Next, if we
perform an orhogonal transformation $O\in O(2;\mathbb{R})$ (which is also
symplectic), then from Lemma \ref{LemmaOrthosymplectic} we conclude that we
can make 
\begin{equation}
Wg(z)=Wg^{\prime }\left( Oz\right) =\sum_{n,m=0}^{N}c_{n}\overline{c_{m}}%
W(h_{n},h_{m})(z)~,  \label{eqProofLineSegment2A}
\end{equation}%
with $c_{n}=d_{n}e^{i\alpha n}$, with some $\alpha \in \left[0, 2 \pi \right]
$, vanish on the line segment 
\begin{equation*}
\left\{ (x,0)\in \mathbb{R}^{2}:~0 \leq x\leq \gamma \right\}~,
\end{equation*}
for some $\gamma >0 $. We infer: 
\begin{equation}
\begin{array}{c}
\sum_{n=0}^{N}|c_{n}|^{2}(-1)^{n}L_{n}(\xi ^{2})+ \\ 
\\ 
+\sum_{n=0}^{N-1}\sum_{k=1}^{N-n}(-1)^{n}\sqrt{\frac{n!}{(n+k)!}}\xi
^{k}L_{n}^{(k)}(\xi ^{2})2Re(c_{n}\overline{c_{n+k}})=0~,%
\end{array}
\label{eqProofLineSegment3}
\end{equation}%
for all $\xi =\sqrt{\frac{2}{\hbar }}x\in \left[ 0 ,\sqrt{\frac{2}{\hbar }}%
\gamma \right] $. By a simple inspection we conclude that the highest power
of the variable $\xi $ appearing in eq.(\ref{eqProofLineSegment3}) is $\xi
^{2N}$ with coefficient $\frac{(-1)^{N}|c_{N}|^{2}}{N!}$. Consequently $%
c_{N}=0$, and we have a contradiction.

\subsection{Proof of Proposition \protect\ref{PropositionIsolatedZeros}}

\begin{enumerate}
\item[(i)] We start by showing that if $N \leq 2$, then there cannot be any
isolated zeros.

Suppose that $Wf$ does have isolated zeros. Then, after possibly applying a
time-frequency shift and the pseudo-displacement operator, we can assume,
without loss of generality, that the isolated zero is located at the origin
and that $Wf$ is of the form (\ref{eqpseudoDisp31}), with $g \in \mathcal{H}%
_0^N$ and $N \leq 2$. If $N=0$ there are no zeros. If $N=1$ or $N=2$, and $%
z=0$ is an isolated zero, then it must also be a local minimum or a local
maximum of $Wf$. Moreover, it must also be a zero and local maximum or
minimum of the polynomial part. If we write 
\begin{equation}
Wf(z)= \frac{e^{-\frac{|z-z_0|^2}{\hbar}}}{\pi \hbar} \mathcal{P}_N (z)~,
\label{eqIsolatedZeros0A}
\end{equation}
we must have: 
\begin{equation}
\left\{ 
\begin{array}{l}
\mathcal{P}_N (0,0)=0 \\ 
\nabla \mathcal{P}_N (0,0)=(0,0)%
\end{array}
\right.  \label{eqIsolatedZeros0B}
\end{equation}
The first equation yields: 
\begin{equation}
\sum_{n=0}^{N} (-1)^N |c_n|^2=0~.  \label{eqIsolatedZeros0C}
\end{equation}
The second equation in (\ref{eqIsolatedZeros0B}) amounts to: 
\begin{equation}
\sum_{n=0}^{N-1}(-1)^n c_n \overline{c_{n+1}} \sqrt{n+1}=0 ~ .
\label{eqIsolatedZeros0D}
\end{equation}
If $N=1$, this becomes: 
\begin{equation}
|c_0|=|c_1|~, \hspace{0.3 cm} c_0 \overline{c_1}=0~,
\label{eqIsolatedZeros0E}
\end{equation}
which means that $Wf \equiv 0$.

Alternatively, if $N=2$, then: 
\begin{equation}
\left\{ 
\begin{array}{l}
|c_0|^2-|c_1|^2+|c_2|^2=0 \\ 
\\ 
c_0 \overline{c_1}= \sqrt{2} c_1 \overline{c_2}%
\end{array}
\right.  \label{eqIsolatedZeros0F}
\end{equation}
If $c_1=0$, then we conclude from the first equation that $c_0=c_2=0$, and $%
Wf$ vanishes identically. So, if $c_1 \neq 0$, the second equation yields: 
\begin{equation}
|c_0|=\sqrt{2}|c_2|~.  \label{eqIsolatedZeros0G}
\end{equation}
From the first equation in (\ref{eqIsolatedZeros0F}) and the normalization
condition, we obtain: 
\begin{equation}
|c_0|=\frac{1}{\sqrt{3}}~, ~|c_1|=\frac{1}{\sqrt{2}}~, ~|c_2|=\frac{1}{\sqrt{%
6}}~.  \label{eqIsolatedZeros0H}
\end{equation}
Next, we have: 
\begin{equation}
\left\{ 
\begin{array}{l}
\frac{\partial^2 \mathcal{P}}{\partial x^2}(0,0)=\frac{4}{\hbar} \left[%
|c_1|^2-2|c_2|^2 + \sqrt{2} \text{Re} (c_0 \overline{c_2}) \right] \\ 
\\ 
\frac{\partial^2 \mathcal{P}}{\partial p^2}(0,0)=\frac{4}{\hbar} \left[%
|c_1|^2-2|c_2|^2 - \sqrt{2} \text{Re} (c_0 \overline{c_2}) \right] \\ 
\\ 
\frac{\partial^2 \mathcal{P}}{\partial x \partial p}(0,0)=-\frac{4}{\hbar} 
\sqrt{2} \text{Im} (c_0 \overline{c_2})%
\end{array}
\right.  \label{eqIsolatedZeros0I}
\end{equation}
Thus, the determinant of the Hessian matrix $H(x,p)$ is: 
\begin{equation}
\det \left[H(0,0) \right] = \frac{16}{\hbar^2} \left[\left(|c_1|^2 -2
|c_2|^2 \right)^2 -2|c_0 \overline{c_2}|^2 \right]= - \frac{4}{3 \hbar^2} <
0~,  \label{eqIsolatedZeros0J}
\end{equation}
where we used (\ref{eqIsolatedZeros0H}), and we conclude that $(0,0)$ is in
fact a saddle point.

\item[(ii)] Now we show that, for $N \geq 3$, Wigner functions can have
isolated points. Consider the function: 
\begin{equation}
f(x)=\frac{1}{\sqrt{2}}h_0(x)+\frac{1}{\sqrt{2}}h_3(x)~.
\label{eqIsolatedZeros1}
\end{equation}
The associated Wigner distribution is 
\begin{equation}
Wf(x,p)=\frac{e^{-\frac{x^2+p^2}{\hbar}}}{\pi \hbar} \times \mathcal{P}%
(x,p)~,  \label{eqIsolatedZeros2}
\end{equation}
where the polynomial part is: 
\begin{equation}
\mathcal{P}(x,p)= \frac{2}{3 \hbar^3}(x^2+p^2)^3-\frac{3}{\hbar^2}%
(x^2+p^2)^2+\frac{3}{\hbar}(x^2+p^2)+\frac{2}{\hbar \sqrt{3 \hbar}}x^3-\frac{%
6}{\hbar \sqrt{3 \hbar}} xp^2~.  \label{eqIsolatedZeros3}
\end{equation}
Thus, to second order the Taylor series is 
\begin{equation}
\mathcal{P}(x,p) \approx \frac{3}{\hbar}(x^2+p^2)~.  \label{eqIsolatedZeros4}
\end{equation}
Consequently, $(x,p)=(0,0)$ is a zero of $Wf(x,p)$, which is also an
isolated local minimum.
\end{enumerate}

\section*{Acknowledgements}

The authors would like to thank Pedro Costa Dias for useful insights
concerning the proof of Theorem \ref{TheoremConvolution} and zeros of
Laguerre polynomials.

The authors would also like to thank Professors Kathy Driver and Michael
Filaseta for useful discussions on the zeros of Laguerre polynomials.

The authors would like to thank Professor Stephan De Bi\`{e}vre for
organising the QuiDiQua conference \textit{`Quasiprobability Distributions
in Quantum Mechanics and Quantum Information'} (November 2023, Lille,
France), where some ideas were discussed. Discussions around this topic
started in the event `\textit{Quantum Harmonic Analysis and Symplectic
Geometr}y', (April 2018, Strobl, Austria), in celebration of Maurice de
Gosson's seventieth birthday, to whom this paper is dedicated.

L. D. A. was supported by the Austrian Science Fund (FWF), \newline
10.55776/PAT8205923.

U.C. acknowledges funding from the European Union's Horizon Europe Framework
Programme (EIC Pathfinder Challenge project Veriqub) under Grant Agreement
No.~101114899.

The work of N.C.D and J.N.P. was supported by the Funda\c{c}\~ao para a
Ci\^encia e Tecnologia (FCT), Portugal, via the research center GFM project
UID/00208/2025.

\pagebreak

************************************************************

\textbf{Author's addresses:}

\begin{itemize}
\item \textbf{Lu\'{\i}s Daniel Abreu:} Institute of Mathematics, University
of Vienna, Nordbergstrasse 15, 1090 Vienna, Austria.

e-mail address: abreuluisdaniel@gmail.com

\item \textbf{Ulysse Chabaud:} DIENS, \'Ecole Normale Sup\'erieure, PSL
University, CNRS, INRIA, 45 rue d'Ulm, Paris 75005, France.

e-mail address: ulysse.chabaud@inria.fr

\item \textbf{Nuno Costa Dias and Jo\~ao Nuno Prata: } ISCTE - Lisbon
University Institute, Av. das For\c{c}as Armadas, 1649-026, Lisboa, Portugal,%
\newline
and Grupo de F\'{\i}sica Matem\'{a}tica, Departamento de Matem\'atica,
Instituto Superior T\'ecnico, Universidade de Lisboa, Av. Rovisco Pais,
1049-001 Lisboa, Portugal.

e-mail addresses: nunocdias1@gmail.com, joao.prata@mail.telepac.pt
\end{itemize}

\end{document}